\documentclass[aps,prd,superscriptaddress]{revtex4}
\usepackage{amsmath}
\usepackage{amssymb}
\usepackage{amsfonts}
\usepackage{graphicx}
\usepackage{bm}
\usepackage{bm}
\usepackage{cancel}
\usepackage{times,float}
\usepackage{graphicx}
\usepackage[usenames,dvipsnames,svgnames]{xcolor}
\usepackage{hyperref}
\hypersetup{colorlinks=true, linkcolor=NavyBlue, citecolor=PineGreen,urlcolor=cyan}
\usepackage{multirow}
\usepackage{soul}
\usepackage{ulem}
\usepackage{color}

\usepackage{soul}

\usepackage{ulem}
\usepackage{color}

\newcommand{\beq}{\begin{equation}}
\newcommand{\eeq}{\end{equation}}
\newcommand{\barr}{\begin{eqnarray}}
\newcommand{\earr}{\end{eqnarray}}
\newcommand{\mbf}{\mathbf}
\newcommand{\gv}[1]{\ensuremath{\mbox{\boldmath$ #1 $}}}

\usepackage[english]{babel}
\usepackage{empheq}
\usepackage{epsfig}

\newcommand{\mfb}{\mathfrak{b}}

\newcommand{\mfc}{\mathfrak{c}}

\newcommand{\mfr}{\mathfrak{r}}

\newcommand{\mfT}{\mathfrak{T}}

\newcommand{\mft}{\mathfrak{t}}

\begin{document}

\title{Momentum non-conservation in a scalar quantum field theory with a
planar $\theta$ interface }
\author{Daniel G. Vel\'azquez}
\email{danielgv@ciencas.unam.mx}
\affiliation{Facultad de Ciencias, Universidad Nacional Aut\'onoma de M\'exico, \\
04510 M\'exico, Distrito Federal, Mexico\looseness=-1}
\author{R. Mart\'\i nez von Dossow }
\email{ricardo.martinez@correo.nucleares.unam.mx}
\affiliation{Instituto de Ciencias Nucleares, Universidad Nacional Aut\'onoma de M\'exico, \\
04510 M\'exico, Distrito Federal, Mexico\looseness=-1}
\author{Luis F. Urrutia}
\email{urrutia@nucleares.unam.mx}
\affiliation{Instituto de Ciencias Nucleares, Universidad Nacional Aut\'onoma de M\'exico, \\
04510 M\'exico, Distrito Federal, Mexico\looseness=-1}

\begin{abstract}
 
Motivated by the recent interest aroused by non-dynamical axionic electrodynamics in the context of topological insulators and Weyl semimetals, we discuss a simple model of the magnetoelectric effect in terms of a $\theta$-scalar field that interacts through a delta-like potential located at a planar interface. Thus, in the bulk regions the field is constructed by standard free waves with the absence of evanescent components. These waves have to be combined into linear superposition to account for the boundary conditions at the interface in order to yield  the corresponding normal modes. Our aim is twofold: first we quantize the $\theta$-scalar field using the normal modes in the canonical approach and then we look for applications 
 emphasizing  the effect of  momentum non-conservation  due to the presence of the interface. To this end we calculate the decay of a standard scalar particle into two $\theta$-scalar particles showing  the opening of new decay channels.  As a second application we deal with
 the two body  scattering of standard charged scalar particles mediated by a $\theta$-scalar particle, focusing on the momentum non-conserving contribution of the scattering amplitude ${\cal M}^{NC}$. We define a generalization of the usual cross section in order to quantify the emergence of these events. We also  study the allowed kinematical region for momentum non-conservation as well as the position of the poles of the amplitude ${\cal M}^{NC}$. Finally, the  ratio of the magnitudes between ${\cal M}^{NC}$ and the momentum conserving amplitude is discussed in the appropriate region of momentum space.      
\end{abstract}

\maketitle

\section{Introduction}

The study of quantum field theories in the presence of interfaces gained  relevance in quantum electrodynamics, when Carniglia and Mandel \cite{carniglia} introduced a technique for quantizing evanescent waves (which result when a beam of light travels from a medium with high refractive index to a medium with low refractive index) in a system where each  half-space, separated by the $z=0$ plane, is filled with two different dielectric. The procedure  is based in finding the normal modes of the equations of motion that included the contribution of the dielectric medium. Such normal modes, which now replace the standard plane wave basis, are wave triplets formed by an incident, a reflected and a transmitted wave and are called the ingoing modes. Moreover, these modes form a complete and orthogonal set which  allow  to quantize the electromagnetic field in terms of modified creation and annihilation operators. 
Further progress was made in Ref. \cite{glauber}  who
showed that there are changes in the spontaneous emission rates for electric and magnetic dipole transitions of excited atoms within or near a dielectric media.  They compared  the vacuum quantization scheme with that of the system including  the dielectric and proved that the coefficients connecting the creation and annihilation operators of both methods can be interpreted as scattering amplitudes. The  wave triplet quantization scheme was successfully used in Ref.  
\cite{janowicz}, which studied the radiation of a harmonic oscillator in the presence of the same vacuum-dielectric planar interface of Refs. \cite{carniglia,glauber} and evaluated  the radiative frequency shifts due to the interface. Later, it was observed in Ref. \cite{inoue} that although the basis proposed in Ref. \cite{carniglia} includes interface effects and can describe sources, it was necessary to introduce a new basis (the outgoing modes) to be able to describe sinks. These new  modes are obtained by making  time-reversal and spatial-rotation transformations  on the original wave triplet basis. Further use of the the mode triplet basis  was investigated   
 to determine the energy-level shift of a ground-state atom in front of a nondispersive dielectric half-space
\cite{eberlein1}, and  
the self-energy of a free electron in the presence of a flat dielectric surface at the level of one loop \cite{eberlein2}. Recent progress in the study of the quantization of systems with interfaces
which include evanescent waves is reported in Ref. \cite{OCKL} and references therein.  

Nevertheless, permittivity and permeability are not the only parameters that characterize a dielectric. A huge class of important materials, called magnetoelectrics,  incorporate the magnetoelectric susceptibility (MES) $\Theta$ that modifies the constitutive relations as
$
\bm{D}=\epsilon \bm{E} + \Theta \bm{B}, \bm{H}=\frac{1}{\mu} \bm{B}- \Theta \bm{E},
$
giving rise to the so called magnetoelectric effect. This effect produces a coupling between the electrical and magnetic properties of a material allowing  magnetic fields to generate polarization and electric fields to give rise to magnetization. The prediction of this property in antiferromagnetic materials is credited to Landau and Lifshitz \cite{LL}. The effect was also predicted in 1959 by Dzyaloshinskii \cite{DZYA}, and in 1960 it was confirmed experimentally  in ${\rm Cr}_2{\rm O}_3$, which is indeed an antiferromagnetic material \cite{ASTROV}. The first investigations on the magnetoelectric effect are condensed in Ref. \cite{LODELL}. A recent update of this study, including new methods for designing magnetoelectric materials, new experimental techniques, and theoretical concepts for understanding the  magnetoelectric behavior is  reported in Ref. \cite{fiebig}.

Unfortunately, the magnetoelectric response is very much suppressed with the respect to the normal effects due to $\epsilon$ and $\mu$. This motivates the search for large magnetoelectric couplings which  continues mainly in multiferroic compounds, 
which are  characterized by presenting spontaneous and simultaneous polarization and magnetization \cite{ferrites}. Other phases giving rise to the magnetoelectric effect are found in  magnetically and/or electrically polarizable media \cite{NSPALDING,WEERENSTEIN}. This  range of materials has being  recently extended with the discovery of time reversal invariant  topological insulators  (TIs).  Microscopically, these materials are insulators in the bulk but have conducting surface states  stemming  from their  peculiar band structure \cite{PSCR1,PSCR2}. Topological phenomena in condensed matter goes back to Ref. \cite{twentytwo}, where the conductivity of the quantum Hall effect \cite{fourteen} was identified with the first Chern number of the Berry curvature in the reciprocal space. The existence of TIs in two-dimensional ${\rm HgTe}$ quantum wells was predicted in Ref. \cite{one} and some time later confirmed  experimentally \cite{fifteen}. Subsequently, the phenomenon was generalized to three-dimensional systems with  theoretical predictions in Refs. \cite{FUKANE,FUMELE,MB, monopole3,ROY,33}, followed by an  experimental confirmation in Ref. \cite{ten}.

Work  searching for the generation  of  measurable magnetic fields via the magnetoelectric effect includes the study of  point-like electrical sources \cite{science,DYSIA2019} and metallic spheres of finite radius \cite{MAZE} placed  in front of a planar magnetoelectric medium that occupies the half-space $z>0$. Also the effects of an electric dipolar source in the form of  a semispherical capacitor  surrounded by a spherical topologically insultating shell have been investigated \cite{daniel}. The electric field manipulation of the magnetoelectric effect is an exciting new area of research with the potential to impact magnetic data storage, spintronics and high frequency magnetic devices, leading to smaller and more energy efficient devices. 
Recent advances in the manufacture of electrically manipulable magnetoelectric materials \cite{Matsukura,LOTTERMOSER} and new developments in coating techniques for conductors \cite{COATING}, could give viability to new configurations that give rise to stronger and controllable magnetic fields.

The properties of a conventional insulator are determined by its permittivity $\epsilon$ and its permeability $\mu$. The Maxwell equations that describe the behavior of such materials can be derived from the Lagrangian $\mathcal{L}_{\rm em}=(1/8\pi)\left(\varepsilon\mathbf{E}^2-(1/\mu)\mathbf{B}^2 \right)-\rho \Phi +(1/c)\mathbf{J}\cdot \mathbf{A}$,  once the fields are expressed in terms of the electromagnetic potentials ${\mathbf A}, \Phi$,  where $\rho$ and $ \mathbf{J}$ are the charge and current densities and $c$ is the speed of light. To include the magnetoelectric effect, one must add the term $\mathcal{L}_{\Theta}=\frac{\alpha}{4 \pi^2}\Theta(x) \mathbf{E}\cdot\mathbf{B}$,
where $\alpha$ is the fine structure constant and $\Theta(t, \mathbf{x})$ is the MES.  The system defined by ${\cal L}_{\rm em}+ {\cal L}_\Theta$  is commonly referred to as axion electrodynamics \cite{sikivie} or Carroll-Field-Jackiw electrodynamics \cite{CFJ}. When $\Theta$ is constant the term $\mathcal{L}_{\Theta}$ is a total derivative which does not modifies Maxwell equations. Nevertheless,  $\mathcal{L}_{\Theta}$ has physical consequences in systems where $\Theta$  is a function of the space-time coordinates \cite{0810}.  
It is important to bear in mind that the effective equations emerging  from the addition of the  term $\mathcal{L}_{\Theta}$ can describe diverse physical phenomena according to the different choices of $\Theta(t, \mathbf{x})$. For instance: $\Theta$ real and  arbitrary yields the electromagnetic response of  general magnetoelectrics, $\Theta=0, \pi$ describes TIs  \cite{monopole3}, $\Theta \in \mathbb{C}$ gives  the electrodynamics of metamaterials  \cite{magnetoelectric,magnetoelectric12} and $\Theta(\mathbf{x},t)=2\mathbf{b}\cdot \mathbf{x}-2b_0t$ provides the response of Weyl semi-metals \cite{magnetoelectric,magnetoelectric13,4}. It is interesting to observe  that in high energy  physics the term $\mathcal{L}_{\Theta}$   describes the interaction with the electromagnetic field 
of an hypothetical dynamical axionic field $\Theta(t, \mathbf{x})$  \cite{3,33}, which remains good a candidate for the particle that constitute dark matter \cite{darkmatter}. 
The simplest way to observe  the magnetoelectric effect is
to place adjacent media with different constant values values $\Theta_1$ and $\Theta_2$, such that the interface between them 
 provides the necessary $\partial \Theta\neq 0$. Even more, in order to minimize the usually dominant electromagnetic effects one can choose the materials such that they are non-magnetic and have  almost identical permittivities.

The electromagnetic phenomena described above can be viewed as particular cases of field theories  where  space in divided at least into two regions, characterized  by a set of discontinuous parameters across the  interface. The necessity of a quantum version in these cases has motivated us to review its construction in the simplest possible setting incorporating a tractable analogy of the magnetoelectric coupling.
Taking as inspiration  the case of axion electrodynamics  we consider a system which has  only the analogous of a magnetoelectric interface, i.e. such  that on each side of the interface there are media that have a different value of a constant parameter  playing  the role of the  MES. 

We look at  the simplest case  of a scalar field in the presence of  a planar interface at $z=0$ and study the consequences that this has on some physical processes. 
To this end we start from  the Lagrangian of a real massive scalar field  $\Phi$ to which we add a contribution proportional to  $\theta^\mu \partial_\mu \Phi^2$, that is a total derivative when $\theta^{\mu}$ is constant everywhere. This constitutes a simple model that mimics the ${\cal L}_\Theta$ coupling of axion electrodynamics. The vector $\theta^\mu$ is chosen  to describe a system with a planar interface at $z=0$.  Along the work we pay attention to one important feature 
of our system, which  is the non-conservation of momentum along the $z$-axis due to the presence of the interface and highlight its physical consequences. 

A large number of technical details are summarized in the Supplementary Materials \cite{SuppMaterials} and only the results are included in the main text. 

The work is organized as follows. In section \ref{IIA} we define our model in terms of a scalar field  $\Phi$, to be called the $\theta$-scalar field,   obtaining the equations of motion together with the boundary condition at the interface. Section \ref{IIB} discusses the normal  modes of the system which result in triplets of plane waves due to the boundary conditions at the interface. Both ingoing and outgoing basis are considered, their  orthogonality and completeness are established and the relation between them  is also determined. 
The canonical quantization of $\Phi$ in both basis is done in section \ref{HAM},  where we obtain the respective Feynman propagators.
In section \ref{III} we deal with the first application  of this  system by discussing    the decay of a standard scalar particle (one which  does not see the interface) into a pair of $\theta$-scalar $\Phi$ particles. We calculate the mean life of the initial particle  showing the effects of momentum non-conservation. 
In section \ref{IV} we consider  a second application  
consisting on the scattering of two equally charged  standard scalar particles  mediated by the   $\theta$-scalar particle. Since momentum is not conserved, we study the kinematics  in the initial center of mass frame in detail identifying the undetermined variables of the final state. Next, we obtain  the scattering  amplitude for the process and focus  in the momentum-non-conserving contribution. We calculate the transition probability per unit time for these events and define the analogous of a differential cross section which can be measurable. We identify the region in momentum space where momentum violation is allowed and also look at the poles of the scattering amplitude determining their location in  the allowed zone. We present polar plots
showing  this features for the simplifying case when the ingoing particles  impinge perpendicularly to the interface, for some selected choices of the particle parameters. Finally we calculate  the absolute value of the ratio between the non-conserving scattering amplitude and  the standard one in the coexisting  kinematic  region, for some particular cases.  We close in section \ref{V} with a summary and conclusions.

\section{The  model}

\subsection{Lagrangian and  equations of motion}

\label{IIA}

\begin{figure}[h!]
\centering
\includegraphics[scale=0.45]{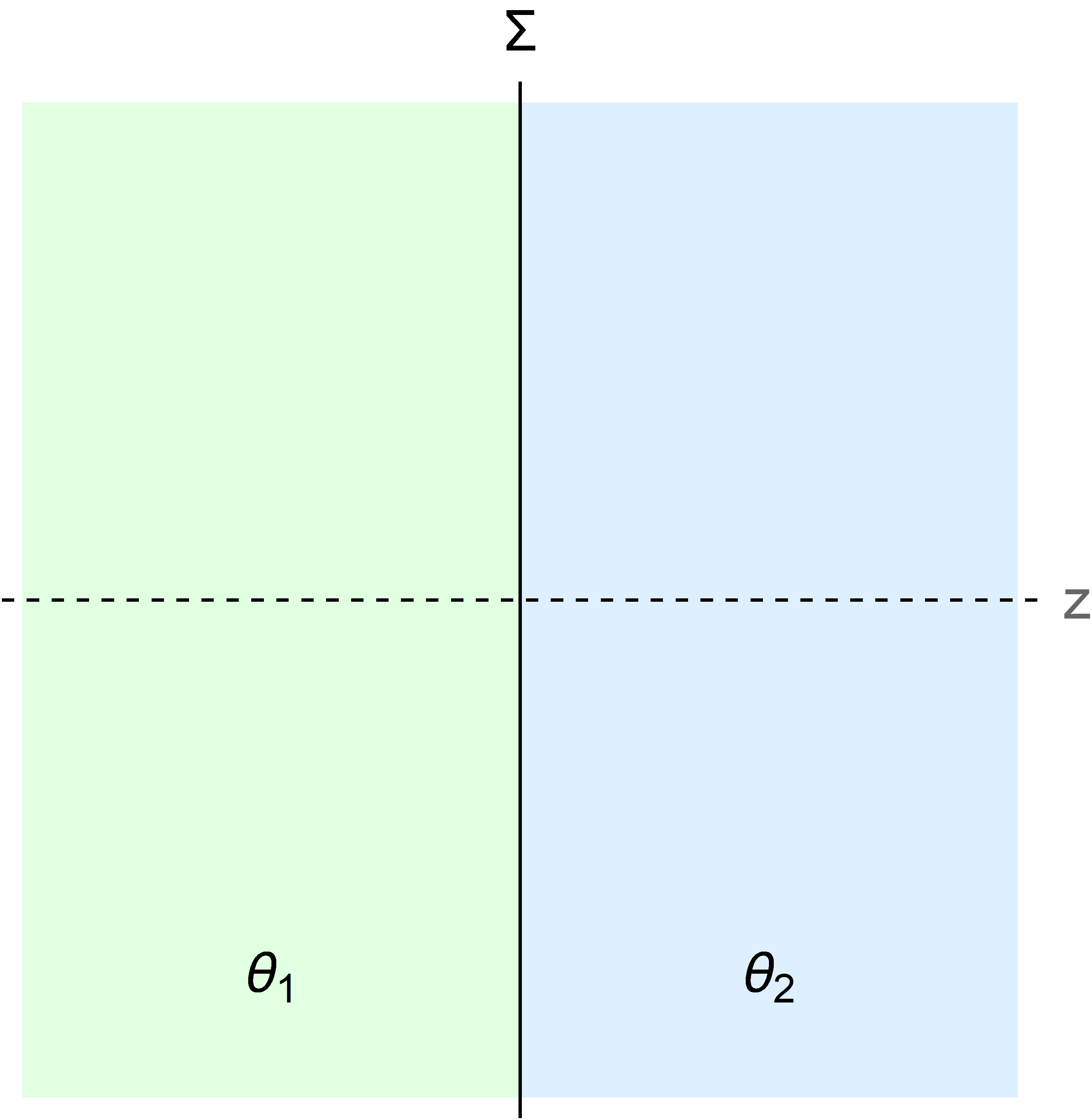}
\caption{\textit{Planar interface $\Sigma$ dividing two regions ($z<0$ and $z>0$) with different values of the $\theta$ parameter.}}
\label{system}
\end{figure}

In analogy with the Lagrangian density describing non dynamical axion electrodynamics ${\cal L}_{\rm em}+ {\cal L}_\Theta$, let us consider the following addition to the standard massive scalar field
\beq
{\cal L}= \frac{1}{2} \partial_\mu \Phi \partial^\mu \Phi-\frac{m^2}{2}\Phi^2-\theta^\alpha(x) \Phi \partial_\alpha \Phi,
\label{LAG1}
\eeq
yielding the  equation of motion
\beq
\Big(\partial^2+ m^2 -\partial_\alpha \theta^\alpha(x) \Big)\Phi=0,
\label{EQM}
\eeq
of the $\theta$-scalar field  $\Phi$.
For a constant $\theta^\alpha$ the last term in the right hand side of  Eq. (\ref{LAG1}) is a total derivative which does not modify the dynamics of the free particle, as can be seen  in Eq. (\ref{EQM}). An interesting and non-trivial situation arises when $\theta^\alpha$ is a time independent piecewise constant function of the coordinates, which we take as
\beq
\theta^\alpha(\gv{x})= \delta^\alpha_3 \Big( \theta_1 H(-z)+ \theta_2 H(z)\Big), 
\label{THETAALPHA}
\eeq 
where $\theta_1$ and $\theta_2$ are constant real numbers and $H(z)$ is the Heaviside function. This choice yields
\beq
\partial_\alpha \theta^\alpha(x) ={\tilde \theta} \delta(z), \qquad {\tilde \theta}=\theta_2-\theta_1,
\eeq
together with
\beq
\Big(\partial^2+ m^2 -{\tilde \theta}\delta(z) \Big)\Phi=0.
\label{EQM1}
\eeq
In this way  we obtain a simple model of a plane interface $\Sigma$ located at $z=0$, which modifies the particle propagation  by  breaking translational invariance along the $z$-axis yielding  momentum non-conservation in that direction. This setup is shown in the Fig. \ref{system}. Let us observe that for $z\neq 0$ we recover the standard free equations of motion for the massive spin zero particle.   

The presence of the interface is dealt with as usual by finding the solutions in the regions $z \neq 0$ and subsequently matching them at $z=0$ according to the boundary conditions at the interface. We observe that in the bulk regions we have a free particle meaning that the components of the momentum are always real. Thus our $\theta$-model does not provide the realization of evanescent waves at the interface.  
Separation of variables yields the linearly independent solutions of Eq. (\ref{EQM1}) as 
\beq
\phi(x)=e^{-iE_{\mbf k} t}e^{i{\mbf k}_\perp \cdot {\mbf x}_\perp}\phi^k(z),
\label{SEPVAR}
\eeq
together with its complex conjugate. Here ${\mbf x}_\perp=(x,y)$,  $\, {\mbf k}_\perp$ denotes the momentum in the direction perpendicular to the $z$-axis (i.e. parallel to the interface) and $k$ is a positive number labeling   the normal modes.  The function  $\phi^k(z)$ satisfies the equation
\beq
\Big(\frac{\partial^2}{\partial z^2}+ E_{\mbf k}^2 -{\mbf k}_\perp^2 -m^2 +{\tilde \theta}\delta(z)  \Big)\phi^k(z)=0.
\label{EQM2}
\eeq
 and will be constructed as  a linear combination of $e^{\pm i k z}$ with  
 $ k=+\sqrt{E^2_{\mbf k}-{\mbf k}_\perp^2 - m^2} \geq 0$, as dictated  by the solutions of Eq. (\ref{EQM2}) outside the interface. This linear combination arises from the boundary conditions at the interface
 which require  the continuity of $\phi^k(z)$ together with the discontinuity in the derivative of $\phi^k(z)$ provided by the integration of Eq.(\ref{EQM2}) around $z=0$. In other words we 
 should fulfill 
 \beq
 \phi_1(z=0_-)=\phi_2(z=0_+)=\phi(z=0),  \qquad  
 \Big(\frac{\partial{\phi_1}}{\partial z}\Big)_{z=0_-}  -
 \Big(\frac{\partial{\phi_2}}{\partial z}\Big)_{z=0_+}= {\tilde \theta} \phi(z=0),
 \label{BC}
 \eeq
in standard notation. From now on the subindices 1 and 2  denote functions living on $z<0$ and $z>0$, respectively.

\subsection{Normal modes:  ingoing (incident)  and  outgoing (detector) modes}

\label{IIB}

\begin{figure}[htb!]
\includegraphics[scale=0.5]{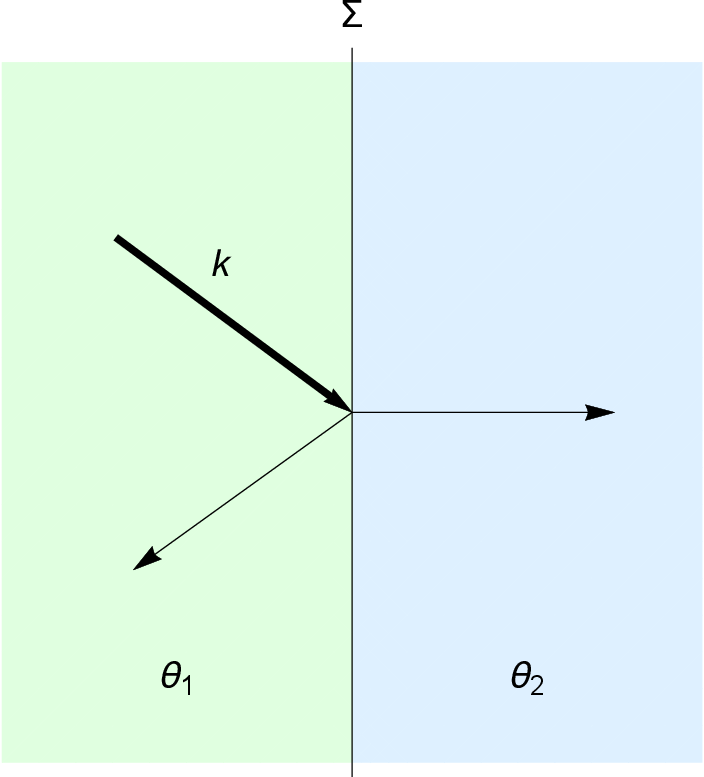} \hspace{1cm}
\includegraphics[scale=0.5]{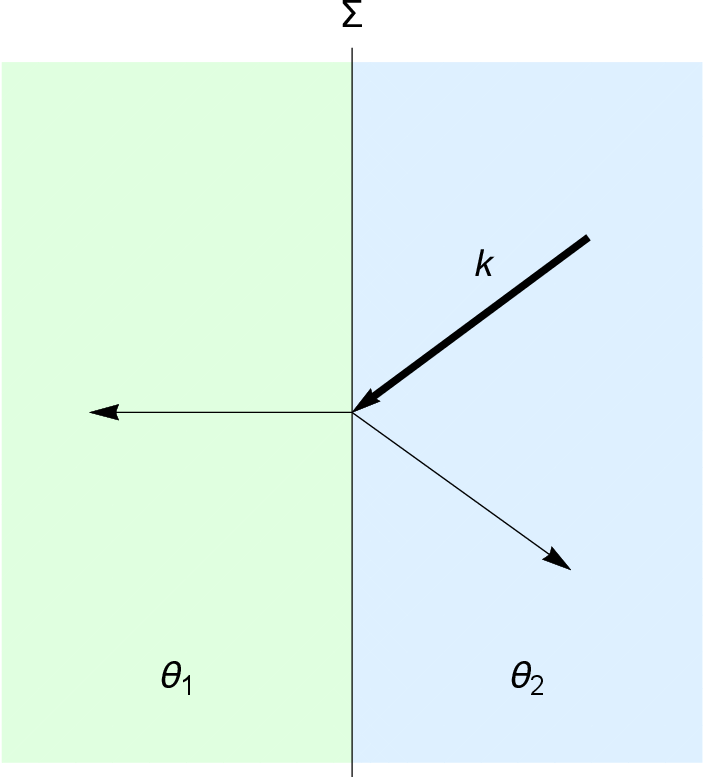}
\caption{\textit{Left panel: the ingoing left mode. Right panel: the ingoing right mode. Both modes include contributions of incident, reflected, and transmitted plane waves.}}
\label{ingoing}
\end{figure}

The first step in the quantization of the proposed model is to find the normal modes associated to Eq. (\ref{EQM1}) in order to describe the propagation non-perturbatively in ${\tilde \theta}$. Due to the interface, a wave impinging on it will be accompanied by reflected and transmitted waves as depicted in Fig \ref{ingoing}. Since the coupling to the interface  is independent of the coordinates perpendicular to the $z$-axis, the momentum ${\mathbf k}_\perp$ parallel to the interface is conserved.

\subsubsection{ The ingoing (incident) modes}

We choose to label the ingoing  normal modes as left (L) and right (R) modes, according to the incident wave  impinging on the interface approaches  from the left or from the right, respectively.  This is emphasized  with the drawing in bold of the vector ${\mbf k}=({\mbf k}_\perp, k_z)$  in the  panels of Fig. \ref{ingoing}. To this end we restrict the label $k$ of these wave functions to $k\geq 0$.

We take the ingoing normal modes to be the functions
\beq
\nu_S^{{\mbf K}}({\mbf x})= e^{i {\mbf k}_\perp\cdot {\mbf x}_\perp} \phi^k_S(z),
\label{INNM}
\eeq
where $S \in \{L, R\}$ and  $\phi^k_S(z)$ satisfies the Eq. (\ref{EQM2}) together with the boundary conditions (\ref{BC}). At this level ${\mbf K}$ is just the  label $({\mbf k}_\perp, k)$. 
The contribution to the normal modes of the plane waves in the perpendicular direction is trivial so that we emphasize the non-trivial dependence upon $\phi_S^k(z)$. In an abuse of language  we refer to the normal modes when dealing only with the $z$-dependent contribution.
 
The construction of the functions  $\phi^k_S(z)$  start by taking  the standard linear combinations of the free plane waves
$e^{\pm i k z}$ and writing  
\barr
&& \phi^k_{1L}= e^{ikz}+P_{Lk}e^{-ikz}, \qquad \phi^k_{2L}= Q_{Lk}e^{ikz} \notag \\
&& \phi^k_{1R}=Q_{Rk}e^{-ikz}, \qquad \phi^k_{2R}=e^{-ikz}+P_{Rk}
e^{ikz},
\label{NM}
\earr
where $P_{Rk}, P_{Lk}$ are the reflected  amplitudes and $Q_{Rk}, Q_{Lk}$ are the transmitted  amplitudes. The boundary conditions (\ref{BC}) yield
\beq
P_{Lk}=P_{Rk}=P_k=-\frac{{\tilde \theta}}{2ik + {\tilde \theta}},
\qquad  Q_{Lk}=Q_{Rk}=Q_k=\frac{2ik}{2ik + {\tilde \theta}},   \qquad Q_k=1+P_k.
\label{DEFPQ}
\eeq
We can verify that $|P_k|^2+|Q_k|^2 =1$, as expected for reflection and transmission coefficients. The relations
\beq
P_k^*=P_{-k}, \qquad Q_k^*=Q_{-k}, \qquad P_{-k} Q_k + P_k Q_{-k}=0,
\eeq
will be useful in the following. From (\ref{NM}) and (\ref{DEFPQ}) the normal  modes are
\barr
&&\phi_L^k(z)=H(-z)(e^{ikz} +P_k e^{-ikz})+H(z)Q_ke^{ikz}
=e^{ikz} +P_k e^{ik|z|}, \label{LMODE}\\
&& \phi_R^k(z)=H(-z)Q_k e^{-ikz}+H(z)(e^{-ikz} +P_k e^{ikz})=
 e^{-ikz}+P_k e^{ik|z|},
 \label{RMODE}
\earr
which satisfy  $\phi^{k *}_{L,R}(z)=\phi_{L,R}^{-k}(z)$ and $\phi^k_{L,R}(z)=\phi_{R,L}^{k}(-z)$.

Our next step is to obtain the scalar product of different normal modes.  The scalar product is
\beq
\langle f|g\rangle =\int_{-\infty}^{+ \infty} d^3x  f^*({\mbf x}) g({\mbf x}),
\label{INNERP}
\eeq
which reduces to
\beq
\langle \nu_S^{{\mbf K}}| \nu_{S'}^{{\mbf K'}} \rangle =(2\pi)^2 \delta^2
({\mbf k}_\perp -{\mbf k}'_\perp)\int_{-\infty}^{+\infty} dz (\phi^k_S)^*(z) \phi^{k'}_{S'}(z) \equiv (2\pi)^2 \delta^2
({\mbf k}_\perp -{\mbf k}'_\perp) \, \langle{\phi_S^{k}}|{\phi_{S'}^{k'}}\rangle.
\label{SCALARP}
\eeq 
The calculations to obtain the scalar product in the $z$-space 
are rather involved and we illustrate    one of them  in the Supplementary Materials \cite{SuppMaterials}. Here we only quote the final results for further use.  We find
\barr
&& \langle{\phi_S^{q}}|{\phi_{S'}^{k}}\rangle= (2\pi) \delta(k-q)
\delta_{S S'},\label{product}
\\
&& \langle{\phi_S^{q*}}|{\phi_{S'}^{k}}\rangle=[\delta_{SS'}P_{q}+(1-\delta_{SS'})Q_{q}] (2\pi) \delta(k-q)=[Q_{q}-\delta_{SS'}] (2\pi) \delta(k-q).
\label{other_products}
\earr
where $S,S'\in \{L,R\}$, which show that the normal modes are orthogonal.

It is convenient to introduce the notation
\beq
U^k_{S S'}= \delta_{SS'}P_{k}+(1-\delta_{SS'})Q_{k}= U^k_{S'S},
\label{MATU}
\eeq
which implies  that the matrix $U^k$ is symmetrical. 
In terms of the complete wave functions (\ref{INNM}) the orthogonality relations read
\begin{eqnarray}
&&\langle{\nu_S^{\mbf K}}|{\nu_{S'}^{{\mbf K}'}}\rangle=(2\pi)^3\delta^{(3)}(\mbf{K}-\mbf{K}') \delta_{SS'},
\label{orthogonality1} \\
&&\langle{\nu_S^{{\mbf K}*}}|{\nu_{S'}^{{\mbf K}'}}\rangle= U^k_{S S'}(2\pi)^3\delta(k-k')\delta^{(2)}(\mbf{k}_{\perp}+\mbf{k}'_{\perp }).
\label{orthogonality2}
\end{eqnarray}

After a long calculation summarized in the Supplementary Materials \cite{SuppMaterials} we establish the completeness relation of the  ingoing normal modes
\begin{equation}
\int_0^{\infty}dk\left(\Phi_L^k(z)\Phi_L^{k*}(z')+\Phi_R^k(z)\Phi_R^{k*}(z') \right) = 2\pi \delta(z-z').
\label{COMPLREL}
\end{equation}
Let us emphasize that the above relation is fulfilled provided ${\tilde \theta} < 0$, which determines the correct choice of the  sign  in the coupling  term of the Lagrangian density (\ref{LAG1}). As we will discover in a subsequent section this condition is also required to have a positive definite Hamiltonian for arbitrary configuration of the fields. 

\subsubsection{The outgoing (detector) modes}

\begin{figure}[htb!]
\includegraphics[scale=0.5]{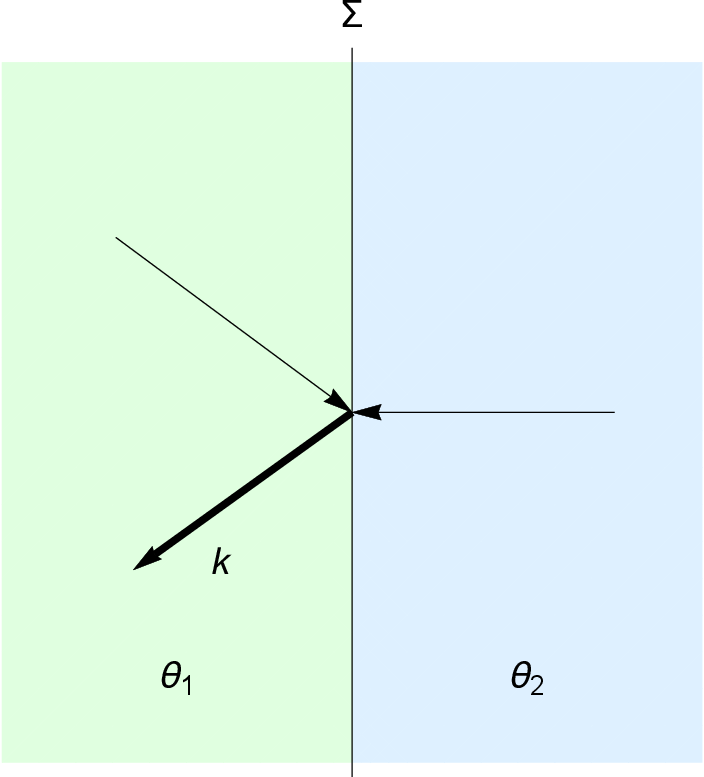} \hspace{1cm}
\includegraphics[scale=0.5]{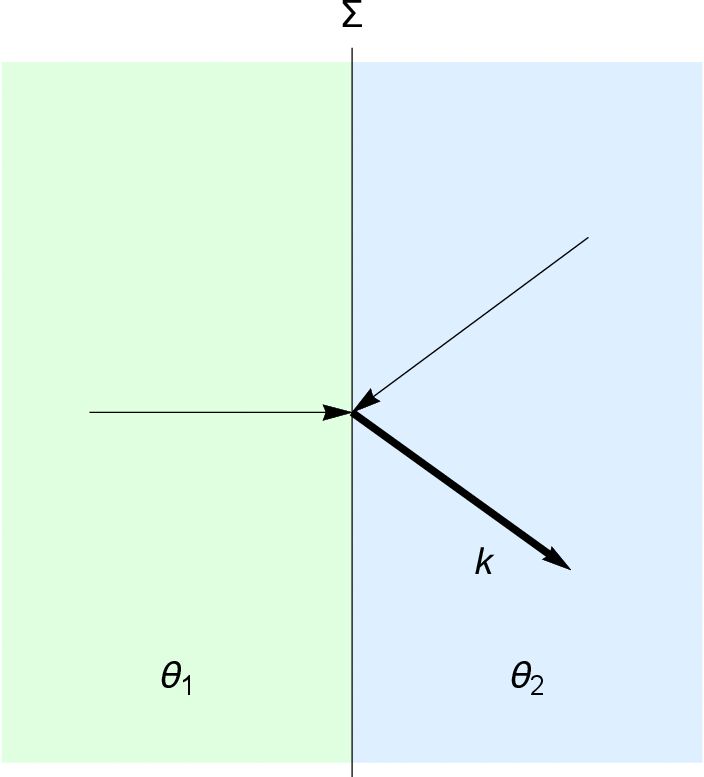}
\caption{\textit{Left panel: the outgoing left mode. Right panel: the outgoing right mode. Both modes  include contributions of outgoing, pseudoreflected and pseudotransmitted plane waves.}}
\label{outgoing}
\end{figure}

So far, we have introduced  a basis of ingoing modes  denoted by $\nu_S^{{\mbf K}}({\mbf x})$  parametrized by $\mbf{K}=({\mbf k}_\perp, k)$. Such basis is adequate for describing particle sources (which impinge at the interface), but not particle sinks (e.g. detectors) associated with outgoing particles. Focusing on the non-trivial dependence of the normal modes in the $z$-component of the momentum, we choose to define the basis of {outgoing} (detector modes)  by means of the substitution ${k} \rightarrow -{k}$ in the functions $\phi_S^{{k}}(z)$ which we already know. Given that  $\phi_S^{{ k} *}({ z})=\phi_S^{-{ k}}({ z})$, the basis of outgoing modes is given by $\{\phi_S^{{ k} *}({ z})\}$ with $k>0$.
Since the outgoing  functions correspond to the complex conjugate of the ingoing functions in the $z$-space, the completeness  relation of the former is automatically satisfied. 
In the new basis, we have an outgoing, a pseudoreflected and a pseudotransmitted wave, as outlined in the  Fig. \eqref{outgoing}. 

In analogy to Eq. (\ref{INNM}) we define the outgoing normal modes as
\beq
{\tilde \nu}_S^{{\mbf K}}({\mbf x})= e^{i {\mbf k}_\perp\cdot {\mbf x}_\perp} \phi^{k *}_S(z).
\label{OUTNM}
\eeq
We warn the reader that we are not taking the complex conjugate of the perpendicular contribution in the above definition.

\subsubsection{The relation between the ingoing and outgoing basis}

Next we inquire about  the relationship between both bases.  To find their connection, we start from the completeness relation:
\begin{equation}
\Phi_S^{k*}(z)=\int dz'\delta(z-z')\Phi_S^{k*}(z')=\int d 
z'\left(\int_0^{\infty}\frac{dk'}{2\pi}\sum_{S'\in\{L,R\}}\Phi_{S'}^{k'}(z)\Phi_{S'}^{k' *}(z')\right)\Phi_S^{k *}(z').
\label{RELMODES}
\end{equation}
This is rewritten in terms of the  known scalar product as
\begin{equation}
\Phi_S^{k *}(z)=\int_0^{\infty}\frac{dk'}{2\pi}\sum_{S'\in\{L,R\}}\left(\int d z'\Phi_{S'}^{k'*}(z')\Phi_S^{k *}(z')\right)\Phi_{S'}^{k'}(z),
\end{equation}
where we know that
\begin{equation}
\int_{-\infty}^{+\infty} dz' \, \Phi_{S'}^{k'}(z')\Phi_{S}^{k}(z')=U^k_{S S'} (2\pi) \delta(k-k'),
\end{equation}
from Eqs. (\ref{other_products}) and (\ref{MATU}).
Thus, taking the complex conjugate yields
\begin{equation}
\Phi_S^{k *}(z)=\sum_{S'\in\{L,R\}}U^{*k}_{S S'} \,\Phi_{S'}^{k}(z), \qquad U^{*k}_{S S'}\equiv (U^k_{S S'})^*,
\label{salientes_entrantes}
\end{equation}
which provides the searched relationship among the basis.
The inverse relation is found by conjugating Eq. \eqref{salientes_entrantes} and yields
\begin{equation}
\Phi_S^{k}(z)=\sum_{S'\in\{L,R\}}U^{k}_{S S'} \,\Phi_{S'}^{*k}(z). 
\label{entrantes-salientes}
\end{equation}
 Eqs. (\ref{salientes_entrantes}) and (\ref{entrantes-salientes}) imply  the following  condition on the symmetrical matrix $U^k_{S S'}$
 \beq
 \sum_{S''\in\{L,R\}}U^{k}_{S S''}\, U^{*k}_{S'' S'}=\delta_{S S'},
 \label{SYMUNIT}
 \eeq
 which is recovered by explicit calculation in the Supplementary Materials \cite{SuppMaterials}. Thus, $U^k_{S S'}$ is a unitary symmetric matrix.

The orthogonality of the outgoing  modes 
\begin{equation}
\langle{\Phi_S^{q*}}|{\Phi_{S'}^{k*}}\rangle=2\pi \delta(k-q)\delta_{SS'},\qquad
\langle{\Phi_S^{q}}|{\Phi_{S'}^{k*}}\rangle=U^{* q}_{S S'} (2\pi) \delta(k-q).
\end{equation}
is a direct consequence of the orthogonality of the ingoing modes.

Recalling Eq. (\ref{INNM}), together with (\ref{entrantes-salientes}), we find the important relation between the ingoing and outgoing modes
\beq
{\nu}_S^{{\mbf K}}({\mbf x})= \sum_{S'\in\{L,R\}}U^{k}_{S S'} \,
{\tilde \nu}_{S'}^{{\mbf K}}({\mbf x}), \qquad {\tilde \nu}_S^{{\mbf K}}({\mbf x})= \sum_{S'\in\{L,R\}}U^{*k}_{S S'} \,
{ \nu}_{S'}^{{\mbf K}}({\mbf x}),
\eeq
which produces
\beq
\sum_{S\in\{L,R\}} {\nu}_S^{{\mbf K}}({\mbf x})
{\nu}_S^{{\mbf K}*}({\mbf x})= \sum_{S\in\{L,R\}} {\tilde \nu}_S^{{\mbf K}}({\mbf x})
{\tilde \nu}_S^{{\mbf K}*}({\mbf x}).
\label{SUMINOUT}
\eeq

\subsubsection{The momentum content of the modes}

So far we have emphasized that ${\mbf K}=({\mbf k}_\perp, k)$ with $k>0$ is just a label for each mode. Here ${\mbf k}_\perp$ is indeed the momentum carried  by the corresponding particle, but the component $k_z$ of the momentum has to be further specified in terms of the label $k>0$ of the mode. This relation is listed   in Table \ref{TABLE1} and it is shown in the Figs. \ref{ingoing} and \ref{outgoing}.

\begin{table}
	\begin{center}
	\begin{tabular}{|c|c|}
	\hline \hline
	$  {\rm Mode} $ & $\qquad k_z \qquad $   \\ \hline \hline
	$ {\rm ingoing \,\, R}  $ &  $\qquad -k \qquad  $      \\ \hline
	${\rm ingoing \,\, L} $ &   $ \qquad + k \qquad$       \\ \hline
	$ {\rm outgoing \,\, R} $ & $\qquad +k \qquad$      \\ \hline 
	$ {\rm outgoing \,\, L}  $ & $ \qquad -k \qquad $        \\  
	\hline \hline
	\end{tabular}
	\caption{For a given label $k >0$ of the mode, we list the values of $k_z$ corresponding to the propagating particle.}
	\label{TABLE1}
	\end{center}
\end{table}

\subsection{The quantization, the Hamiltonian and the propagator}
\label{HAM}

The most general real solution to the modified  Klein-Gordon equation (\ref{EQM}) can be taken  as a linear combination of either the  ingoing normal modes or the outgoing  normal modes. We start with the  former and write
\begin{eqnarray}
\Phi(t,\mbf{x})&=&\sum_{S\in\{L,R\}}\int_{k>0}\frac{d^3 {\mbf K}}{(2\pi)^3}\frac{1}{\sqrt{2E_{\mbf K}}}\left[a_S(\mbf{K})\nu^{\mbf{K}}_S(\mbf{x})e^{-iE_{\mbf K} t}+a_S^\dagger(\mbf{K})\nu_S^{* \mbf{K}}(\mbf{x})e^{iE_{\mbf K} t} \right],
\label{FIELD}
\end{eqnarray}
with $E_{\mbf K}=\sqrt{{\mbf k}_\perp^2+k^2 +m^2}$ and $d^3 {\mbf K}=d^2 {\mbf k}_\perp \, d k$. 

We quantize the field by imposing the following commutation relations
\begin{eqnarray}
\left[a_S(\mbf{K}),a^{\dagger}_{S'}(\mbf{K}') \right]=(2\pi)^3\delta^{(3)}(\mbf{K}-\mbf{K}')\delta_{SS'},\qquad 
\left[a_S(\mbf{K}),a_{S'}(\mbf{K}') \right]=0=\left[a^{\dagger}_S(\mbf{K}),a^{\dagger}_{S'}(\mbf{K}') \right],
\label{CR}
\end{eqnarray}
which reproduce the equal time canonical commutation relations among  the field $\Phi(t, {\mbf x})$ and its canonically  conjugated momentum
$\Pi(t, {\mbf x})=\dot{\Phi}(t, {\mbf x})$,  at different space points. The calculations, which involve the completeness relation (\ref{COMPLREL}), are completely analogous to the standard case so we write the expected final results
\begin{equation}
\left[\Phi(t,\mbf{x}),{\Pi}(t,\mbf{x}')\right]=i\delta^{(3)}(\mbf{x}-\mbf{x}'), \qquad  \left[\Phi(t,\mbf{x}),{\Phi}(t,\mbf{x}')\right]=0, \qquad \left[\Pi(t,\mbf{x}),{\Pi}(t,\mbf{x}')\right]=0. 
\end{equation}
The Fock space of the system generated from the vacuum $
|0 \rangle$, such that $a_S({\mbf K})|0 \rangle=0$, is constructed in the usual way in terms of  the  operators $a_S^\dagger({\mbf K})$ acting on the vacuum.

To identify the Hamiltonian and the momentum operator it is convenient to start from the energy-momentum tensor. To obtain a symmetric energy-momentum tensor we use the Lagrangian density 
\beq
{\cal L}'= \frac{1}{2}\partial_\mu \Phi \partial^\mu \Phi- \frac{m^2}{2} \Phi^2+ \frac{\tilde \theta}{2} \delta(z) \Phi^2
\eeq 
which differs from the original Eq. (\ref{LAG1}) by a total derivative. We obtain
\beq
T^0{}_0=\frac{1}{2}{\dot \Phi}^2+ \frac{1}{2}(\nabla \Phi)^2+
\frac{m^2}{2}\Phi^2+\theta(z) \Phi \partial_z \Phi, \qquad 
T^0{}_i={\dot \Phi} \partial_i \Phi, \label{TMUNU}
\eeq
together with the non-conservation equation
\beq
\partial_\mu T^\mu{}_\nu=-\frac{{\tilde \theta}}{2}\Phi^2 \partial_\nu \delta(z),
\label{NC}
\eeq 
which is consistent with the violation of translational invariance in the direction $z$. In fact, here we have energy conservation and momentum conservation only in the directions  perpendicular to de $z$-axis. Let us observe that the last term of $T^0{}_0$ in Eq. (\ref{TMUNU}) can be integrated by parts in the expression for  the Hamiltonian, yielding the manifestly  negative contribution $-\frac{{\tilde \theta}}{2}\delta(z) \Phi^2$ to the Hamiltonian density. Thus, in order to have a positive definite Hamiltonian for all configurations of the field $\Phi$  we demand ${\tilde \theta}<0$. This is consistent with the condition required for the  completeness relation (\ref{COMPLREL}) obtained in the previous section. As expected, the calculation of the Hamiltonian in terms of the normal modes yields 
\begin{equation}
H = \sum_{S\in\{L,R\}}\int_{k>0}\frac{d^3{\mbf K}}{(2\pi)^3}E_{ \mbf K} \, a_S^{\dagger}(\mbf{K})a_S(\mbf{K}).
\label{Hamiltoniano1}
\end{equation}
 The details are summarized in the 
Supplementary Materials \cite{SuppMaterials}.

From the commutation relations (\ref{CR}) we realize that   $a_S^{\dagger}(\mbf{K})a_S(\mbf{K})$ can be interpreted as a number operator which eigenvalues label  the number of $\{S,\mbf{K}\}$ modes. Then, the usual interpretation of  $a^\dagger_S({\mbf K})$ ($a_S({\mbf K})$) as creation (annihilation) operators for the ingoing normal modes 
$\{S,\mbf{K}\}$ follows. A similar calculation yields the conserved momentum in the perpendicular direction
\beq
{\mbf P}_{\perp }=\int d^3 x \, {\dot \Phi} \nabla_{\perp }\Phi=
\sum_{S\in\{L,R\}}\int_{k >0}\frac{d^3{\mbf K}}{(2\pi)^3}\, { \mbf k}_{\perp} \, a_S^{\dagger}(\mbf{K})a_S(\mbf{K}).
\label{PPERP}
\eeq
The momentum $P_z$ is not conserved and can not be written  as in Eq. (\ref{PPERP}). This can be understood noticing that $\nu_S^{\mbf K}({\mbf x})$, being a linear combination of three plane waves,  is not an eigenfunction of $i \partial_z$, as it is evident from Eqs. (\ref{LMODE}) and (\ref{RMODE}).

We can also write the field (\ref{FIELD}) in terms of the outgoing normal modes. To this end we start from 
\begin{eqnarray}
&& \Phi(t, {\mbf x})=\int_{k >0}\frac{d^3{\mbf K}}{(2\pi)^3}\frac{1}{\sqrt{2E_{\mbf K}}}\left[\left(\sum_{S\in\{L,R\}}a_S(\mbf{K})\phi_S^{k}(z)\right)e^{-iE_{ \mbf K} t+i\mbf{k}_{\perp}\cdot \mbf{x}_{\perp}}+\text{h.c.}\right].
\end{eqnarray}
and use the  Eq. (\ref{entrantes-salientes}) to replace  the  functions 
$\phi_S^k$ in terms of the outgoing modes $\phi_S^{k *}$. This yields
\begin{eqnarray}
&& \hspace{-1.5cm} \sum_{S\in\{L,R\}}a_S(\mbf{K})\phi_S^{k}(z)= \sum_{S, S'\in\{L,R\}}a_S(\mbf{K})U^k_{S S'} \phi_{S'}^{*k}(z)= \sum_{S'\in\{L,R\}}\Big( \sum_{S \in\{L,R\}} U^k_{S' S} \, a_S(\mbf{K})\Big)\phi_{S'}^{*k}(z),
\end{eqnarray}
and suggests defining the annihilation operators for the outgoing modes as
\beq
{\tilde a}_{S'}(\mbf{K})\equiv  \sum_{S \in\{L,R\}} U^k_{S' S} \, a_S(\mbf{K}),
\label{OUTOP}
\eeq
with the corresponding Hermitian conjugates being the creation operators. 
We note that in the limit $\tilde{\theta}\rightarrow 0$, we have
$
{\tilde a}_L(\mbf{K})= a_R(\mbf{K})$ and $ {\tilde a}_R(\mbf{K})=a_L(\mbf{K})$, 
that is, an incoming $L$ mode corresponds to an outgoing $R$ mode, and viceversa. This is the expected behavior.

In the Supplementary Materials \cite{SuppMaterials} we verify that the algebra of the tilde operators is the expected one. Since each of them is a linear combination of the corresponding untilded ones, we only need to check the commutator $[{\tilde a_S}({\mbf K}), {\tilde a^\dagger_{{S}'}}({\mbf K}')]$. Furthermore, we calculate the mixed commutators obtaining
\begin{eqnarray}
&& \left[{\tilde a}(\mbf{K}),a_{S'}(\mbf{K}') \right]=0, \\
&&\left[{\tilde a}_S(\mbf{K}),a_{S'}^{\dagger}(\mbf{K}') \right]
=(2\pi)^3\delta^{(3)}(\mbf{K}-\mbf{K}') U^k_{S S'}.
\end{eqnarray}
In terms of the outgoing (detector) modes the field is 
\begin{equation}
\Phi(t,\mbf{x})=\sum_{S\in\{L,R\}}\int_{k>0}\frac{d^3{\mbf K}}{(2\pi)^3}\frac{1}{\sqrt{2E_{\mbf K}}}\left[{\tilde a}_S(\mbf{K}){\tilde \nu}^{\mbf{K}}_S(\mbf{x})e^{-iE_{\mbf K} t}+\text{h.c.}\right].
\label{FIELD_DET_MODES}
\end{equation}

As shown in the Supplementary Materials \cite{SuppMaterials} the Feynmann propagator can adopt either of the following two forms 
\begin{eqnarray}
{\Delta}_F(x, x')&&=\sum_{S\in\{L,R\}}\int_{k >0} \frac{d^4 {k}}{(2\pi)^4}\frac{i}{k^2-m^2+i\epsilon}e^{-ik^0(x^0-{x'}^0)}{\tilde \nu}_S^\mbf{k}(\mbf{x}){\tilde \nu}_S^{  \mbf{k} * }(\mbf{x'}),\label{feynman_propS}\\
{\Delta}_F(x,y) &&=\sum_{S\in\{L,R\}}\int_{k>0} \frac{d^4k}{(2\pi)^4}\frac{i}{k^2-m^2+i\epsilon}e^{-ik^0(x^0-{x'}^0)}\nu_S^\mbf{k}(\mbf{x})\nu_S^{{\mbf{k}} *}(\mbf{x'}),
\label{feynman_propT}
\end{eqnarray}
when expressed in terms of the outgoing (ingoing) modes, respectively.
The notation is  $k^\mu=(k^0, \mbf{k}_{\perp}, k )$, with
\beq
d^4 k= dk^0 \, d^3 {\mbf K}=dk^0 d{\mathbf k}_\perp dk, \qquad k^2=k_0^2-{\mbf K}^2
= k_0^2-{\mbf k }_\perp^2-k^2.
\label{NOTATION}
\eeq
In the Supplementary Materials \cite{SuppMaterials} we show that the propagator (\ref{feynman_propT}) can also be written as 
\begin{equation}
\Delta_F(x, x')=i\int_{-\infty}^{+\infty} \frac{d^4k}{(2\pi)^4}\frac{e^{-ik^0(x^0-{x'}^0)+i\mbf{k}_{\perp}\cdot(\mbf{x}-\mbf{x'})_{\perp}}}{k^2-m^2+i\epsilon}
\Big( e^{ik(z-z')}+P_{k}e^{ik(|z|+|z'|)}  \Big).
\label{propagador2}
\end{equation}
The same result holds for the representation (\ref{feynman_propS}) in virtue the relation  (\ref{SUMINOUT}). This form has the advantage of exhibiting the momentum violating contribution as the term proportional to $P_k$ inside the  round brackets, as well as the limit
${\tilde \theta}=0, P_k=0$ where the standard form is recovered.

\section{Decay and mean life}

\label{III}

In order to illustrate the consequences of  the momentum non-conservation produced by the interface we consider the decay of a particle described by a  spin cero field $\Psi$ that is not affected by its presence  into two   $\theta$-scalar  particles. The total Lagrangian density is  
\begin{eqnarray}
\mathcal{L}&=&\frac{1}{2}\partial _{\mu }\Phi \partial ^{\mu }\Phi -\frac{m^{2}}{2}\Phi ^{2}-\theta (z)\Phi \partial _{z}\Phi+\frac{1}{2}\partial _{\mu }\Psi \partial ^{\mu }\Psi -\frac{M^{2}}{2}\Psi ^{2}+\lambda \Psi\Phi^2.
\label{LAGPSIPHI}
\end{eqnarray}
 Since we are interested in detecting the outgoing $\theta$-scalar particles, we will express this field  in terms of the detector modes. 
We write the expansion for the field $\Psi$ as
\begin{equation}
\Psi(t,\mbf{x})=\int \frac{d^3p}{(2\pi)^3}\frac{1}{\sqrt{2E_p}}\left[ b(\mbf{p})e^{-ip\cdot x}+\text{h.c.}\right],
\label{PSI}
\end{equation}
and we quantize it in the standard way.
The commutation relations are the usual free-field ones,
\begin{equation}
\left[b(\mbf{p}),b^{\dagger}(\mbf{p}') \right]=(2\pi)^3\delta^{(3)}(\mbf{p}-\mbf{p}'), \qquad \left[b^{\dagger}(\mbf{p}),b^{\dagger}(\mbf{p}') \right]=\left[b(\mbf{p}),b(\mbf{p}') \right]=0.
\label{CRPSI}
\end{equation}
 We add the commutation relations
\begin{equation}
\left[{\tilde a}_S(\mbf{K}),b(\mbf{p}) \right]=\left[{\tilde a}_S^{\dagger}(\mbf{K}),b(\mbf{p}) \right]=\left[{\tilde a}_S(\mbf{K}),b^{\dagger}(\mbf{p}) \right]=\left[{\tilde a}_S^{\dagger}(\mbf{K}),b^{\dagger}(\mbf{p}) \right]=0,
\label{MIXEDCR}
\end{equation}
indicating that $\Phi$ and $\Psi$ are independent fields.

The initial state is merely a particle of the type $\Psi$ with momentum $\mbf{p}$, while the final state is composed of two modes with labels $\{\mbf{K
}_S,\mbf{K}'_{S'}\}$. In terms of creation operators acting on the vacuum of the Fock space this is
\begin{eqnarray}
&&|{i} \rangle \equiv \sqrt{2E_{\mbf p}} \, b^{\dagger}(\mbf{p})\,
 |0 \rangle=|{\mbf{p}}\rangle  \\
&& |{f}\rangle\equiv  \sqrt{2E_{\mbf K}} \, \sqrt{2E_{{\mbf K}'}} \, 
{\tilde a}_S^{\dagger}(\mbf{K})\, {\tilde a}_{S'}^{\dagger}(\mbf{K}')\, |{0}\rangle=|{\mbf{K}_S,\mbf{K}'_{S'}}\rangle=|{\mbf{K}_S}\rangle\otimes|{\mbf{K}'_{S'}}\rangle.
\label{FOCK}
\end{eqnarray}
To first order, the invariant 
amplitude that connects the initial state with the final state  is
\begin{equation}
\mfT=i\lambda\int d^4 x \langle{f}|\mathcal{T}\left\{\Psi(x)\Phi^2(x)\right\}|{i} \rangle,
\label{INVAMP}
\end{equation}
where $\mathcal{T}$ is the time ordering operator. Decomposing the fields in positive and negative frequency parts we have 
\begin{eqnarray}
\Psi^+(x)|{\mbf{p}}\rangle
&=& e^{-iE_{{\mbf p}} x^0} e^{+i{\mbf p}\cdot {\mbf x}}|{0}\rangle, \qquad 
\langle\mbf{K}_S|\Phi^-(x)=\langle 0 |\, e^{iE_{{\mbf K}} x^0}\, {\tilde \nu}_{S}^{{\mbf K} *}(\mbf{x}), 
\label{POSNEGPARTS}
\end{eqnarray}
which yield the following contribution for the decay amplitude  in the center of mass reference frame ($\mbf{p}=0$)
\begin{equation}
\mfT=i 2\lambda \left(\int dx^0 e^{i(E_{\mbf K}+E_{{\mbf K}'}-M)x^0} \right) \left(\int d\mbf{x}_{\perp}e^{-i(\mbf{k}_{\perp}+\mbf{k}'_{\perp})\cdot \mbf{x}_{\perp}} \right)\left(\int d z \, \phi_S^{k}(z)\phi_{S'}^{k'}(z) \right),
\label{AMPCMD}
\end{equation}
where $\mbf{k}_{\perp}=(k_x, k_y)$, $\mbf{k}'_{\perp}=(k'_x,k'_y)$ and $\mbf{x}_{\perp}=(x, y)$. Thus,
\begin{equation}
\mfT=i 2\lambda (2\pi)^3\delta(E_{\mbf K}+E_{\mbf K}'-M)\delta^{(2)}(\mbf{k}_{\perp}+\mbf{k}'_{\perp})\langle{ \Phi_S^{k*}}|{\Phi_{S'}^{k'}}\rangle.
\label{AMPCMD1}
\end{equation}
From the result of Eq. \eqref{other_products} for the last integral we obtain 
\begin{equation}
\mfT=i 2\lambda (2\pi)^4\bigg[\delta_{SS'}P_{k}+(1-\delta_{SS'})Q_{k}\bigg]\delta(M-E_{\mbf K}-E_{\mbf K}')\delta^{(2)}(\mbf{k}_{\perp}+\mbf{k}'_{\perp })
\delta(k-k').
\label{amplitude}
\end{equation}
Note that the delta function $\delta(k-k')$ indicates that $k=k'$. Nonetheless, this does not correspond to momentum conservation, since the labels $k$ are always positive and hence do not specify a given direction. The relation of these labels  with the physical momentum in the direction $z$ of each decaying particle is  determined according to  Table \ref{TABLE1}. Equation \eqref{amplitude} indicates that 
there are various channels for the decaying particle: (i) the decay can go into  a left and a right outgoing modes ($S=L, S'=R$ for example), thus conserving  momentum in the direction $z$, or  (ii) the particle can decay into two left (right) outgoing modes ($S=S'$), going from an initial state with zero momentum to a final state with momentum $-2k$ ($+2k$), respectively,  in the direction $z$.  As already mentioned, this is a consequence of the lack of translation invariance this direction.  As a check of consistency we observe from the definitions of $P_{k}$ and $Q_{k}$ that the case without interface ($\tilde{\theta}=0$) reduces to
\begin{equation}
\mfT=i 2\lambda (2\pi)^4\big[1-\delta_{SS'}\big]\delta(M-E_k-E_k')\delta^{(2)}(\mbf{k}_{\perp}+\mbf{k}'_{\perp})\delta(k-k').
\label{TDECAY}
\end{equation}
This means that the decay amplitude is different from zero only if the outgoing modes are different, so that the particles  have opposite $z$-component of the  momentum with the same absolute value, as expected. 

Going back to the general case, from Eq. \eqref{amplitude}  we extract the amplitude
\begin{equation}
\mathcal{M}=i 2\lambda\bigg[\delta_{SS'}P_{k}+(1-\delta_{SS'})Q_{k}\bigg].
\label{AMPCMD2}
\end{equation} 
Since $\delta_{SS'}(1-\delta_{SS'})=0, \,\,  \forall \, \, S,S'$ and $\delta_{SS'}^2=\delta_{SS'}$,we have
\begin{equation}
|\mathcal{M}|^2=4\lambda^2\bigg[\delta_{SS'}|P_{k}|^2+(1-\delta_{SS'})|Q_{k}|^2 \bigg],
\label{AMP2}
\end{equation}
where $
|P_{k}|^2=\tilde{\theta}^2/[4k^2+\tilde{\theta}^2]$ and $|Q_{k}|^2=4(k)^2/[4k^2+\tilde{\theta}^2]$.

Even though $\delta(k-k')$ is not related to momentum conservation, it provides the combination $\delta^{(2)}(\mbf{k}_{\perp}+\mbf{k}'_{\perp})\delta(k-k')$ similar to $\delta^3({\mbf p}_{\rm out}-{\mbf p}_{\rm in})$ 
 which allows us to take the standard expression for the differential decay rate 
\begin{eqnarray}
d\Gamma_{SS'}
&=& \frac{1}{32\pi^2 M}\frac{d^3 {\mbf K}}{E_\mbf K}\frac{d^3 {\mbf K}'}{E_{{\mbf K}'}}|\mathcal{M}|^2\delta(M-E_{\mbf K}-E_{\mbf K}')\delta^{(2)}(\mbf{k}_{\perp}+\mbf{k}'_{\perp})\delta(k-k').
\end{eqnarray}
The integration over the phase space yields 
\beq
\Gamma_{SS'} =\frac{\lambda^2}{4\pi M^2}\left[\delta_{SS'}\frac{|\tilde \theta|}{2}\sin^{-1}\left(\sqrt{\frac{1}{ 1+\frac{{\tilde \theta}^2}{4a}}} \right)+(1-\delta_{SS'})\left\{\sqrt{a}-\frac{|{\tilde  \theta}|}{2}\sin^{-1}\left(\sqrt{\frac{1}{1+\frac{{\tilde \theta}^2}{4a}}} \right) \right\}\right],
\label{DECRATE}
\eeq
where the details are presented in the Supplementary Materials \cite{SuppMaterials}. Here $a=M^2/4-m^2$ and the presence of $\sqrt{a}$ in Eq. (\ref{DECRATE}) demands $a>0$.  A  consistency check of the above result for $\Gamma_{S S'}$ is that when  $\tilde{\theta}=0$ we recover the well known decay width when there is no  interface 
\beq
\Gamma_0= \frac{\lambda^2}{8 \pi M}\sqrt{1-4m^2/M^2}.
\label{DEC0}
\eeq

Summarizing, there are now three decay channels. One when the outgoing modes are different ($\Gamma_{LR}$), which in the limit $\tilde{\theta}\rightarrow 0$ reduces to $\Gamma_0$,  and two others when the modes are the same ($\Gamma_{RR}=
\Gamma_{LL}$), which arise due to the non-conservation of momentum in the direction of the $z$-axis.  

In the first case we have
\begin{equation}
\begin{split}
\Gamma_{LR}=&\Gamma_0\, \left[1-R\sin^{-1}\left(\frac{1}
{\sqrt{1+
R^2}} \right) \right], \qquad  R=\frac{|\tilde{\theta}|}{\sqrt{M^2-4m^2}}>0.
\label{GAMMALR}
\end{split}
\end{equation}
Notice that the condition $\Gamma_{LR}>0$ sets a new threshold in $M$ for the decay to occur. Since $1/\sqrt{1+R^2} <1 $ the resulting angle $\beta=\sin^{-1}(1/\sqrt{1+R^2})$  is between $0$ and $\pi/2$. Then the condition $(1-R  \pi/2)>0$  guarantees  that the square bracket in (\ref{GAMMALR}) is positive, imposing the stronger constraint
\beq
M^2> 4m^2+ \frac{\pi^2} {4} {\tilde \theta}^2,
\label{NEWTH}
\eeq 
which is equivalent to write $0 <R <2/\pi$.

The remaining cases yield
\begin{equation}
\Gamma_{LL}=\Gamma_{RR}=\Gamma_0 \, \left[R\sin^{-1}\left(\frac{1}{\sqrt{1+R^2}} \right)\right],
\label{DSS}
\end{equation} 
 which only  require the condition
\beq
M^2 > 4m^2.
\eeq
Recapping, when $4m^2 <M^2 < 4m^2+ \pi^2 {\tilde \theta}^2/4 $
only the channels $LL$ and $RR$ are open, with a total decay rate
\beq
\Gamma_{\tilde \theta}=\Gamma_{LL}+ \Gamma_{RR}=
2 \Gamma_0\, \left[R\sin^{-1}\left(\frac{1}{\sqrt{1+R^2}} \right)\right].
\label{TDSS}
\eeq  
Let us emphasize that in this region of $M^2$ the decay proceeds only due to the momentum violating channels.
On the other hand, when $4m^2+ \pi^2 {\tilde \theta}^2/4  < M^2$ the three channels participate with a total decay rate
\beq
\Gamma_T=\Gamma_{\tilde \theta} + \Gamma_{LR}=\Gamma_0
\left[1+R\sin^{-1}\left(\frac{1}
{\sqrt{1+
R^2}} \right) \right]. 
\label{73}
\eeq
The function in square brackets in (\ref{73}) is an increasing function of $R \geq 0$, with value $1$ for $R=0$ and value $1.64$ when $R=2/\pi$. Thus we conclude that $\Gamma_0 < \Gamma_T < 1.64 \,  \Gamma_0$. 

\section{ Momentum non-conservation in a scattering process}

\label{IV}

As another signal of momentum non-conservation in our model we study a  scattering  process mediated by the $\theta$-scalar particle. To simplify the calculation we take its interaction with a complex scalar field $\chi$, describing standard particles with mass ${\tilde M}$ and charges $\pm {\tilde \lambda} $ which are not affected by the interface. The interaction term is  ${\cal L}_{\rm int}={\tilde \lambda} \, \Phi \, \chi^\dagger \, \chi $ and we  consider the  scattering of two incoming $\chi$ particles with the same charge going into two outgoing $\chi$ particles. In this way we only allow the $t$ and $u$ channels, suppressing the $s$ channel due to charge conservation.  Our aim is to give a quantitative  description of  the momentum$-$violating contribution. To set the stage, let us recall the most typical example of a process conserving energy but violating momentum conservation, which  is the scattering of a charge by an external potential (Coulomb scattering by an external electromagnetic field, for example). In our case  the interface plays the role of the external potential providing a $\delta(z)$-type interaction.

\subsection{The kinematics}

\label{IV1}

The normal modes of the $\chi$-particles are standard plane waves and the label ${\mbf p}=\{ {\mbf p}_\perp, p\}$ refers to the usual momentum with $p$ being the $z$-component of ${\mbf p}$. The $\theta$-particle  mediates de interaction and only contributes through the propagator. Nevertheless, since the interaction does not conserve momentum along the $z$-axis,  we need some care when specifying the kinematics of the participating $\chi$ particles.   
The four momenta of the  incident particles are  denoted as
\begin{eqnarray}
&& p_1^\mu=(E_1, {\mbf p}_{1 \perp}, p_1), \qquad  p_2^\mu=(E_2, {\mbf p}_{2 \perp}, p_2),\qquad E_i=\sqrt{{\mbf p}^2_{i \perp}+p_i^2+ {\tilde M}^2 }, \quad i=1,2 ,
\label{INCMOM}
\earr
while those of the outgoing particles follow the same notation with a superindex prime. We locate the incident particles in their center of mass, i.e., $\mbf{p}\equiv\mbf{p}_1=-\mbf{p}_2$, with $p=|{\mbf p}|$. We must bear in mind that, due to the non-conservation of momentum, this frame does not coincide with the center of mass frame of the final momenta. 

 Now we choose a convenient coordinate system according to the following convention: the direction of the incident  center of mass, determined by ${\mbf p}_1$,  forms an angle $\theta$ with the normal ${\hat {\mbf{k}}}$ to the interface, which defines our $z$-axis. The projection of ${\mbf p}_1$ on the interface ($x-y$ plane) defines the $x$-axis of the coordinate system, as shown in Fig. \ref{vector}. 
\begin{figure}
\centering
\includegraphics[width=0.45\textwidth]{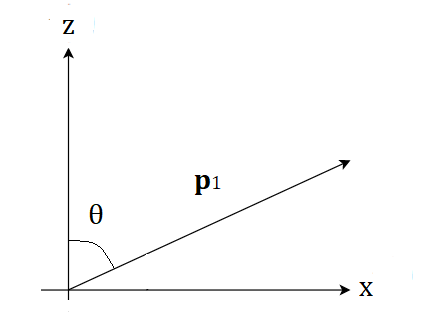}
\caption{Coordinate system for the scattering of two $\chi$ particles mediated by a $\Phi$ particle. The $z$-axis is determined by the  normal to the interface which is the  ($x-y$)-plane. The incident direction ${\mbf p}_1$ of the center of mass forms the angle $\theta$ with the normal to the interface. The projection of ${\mbf p}_1$ on the interface determines de $x$-axis. }
\label{vector}
\end{figure}
In this way
\barr
{\mbf p}_1=(p\sin \theta,0, p \cos\theta), \qquad{\mbf p}_2=(-p\sin \theta,0, -p \cos\theta), \quad E_1=E_2= \sqrt{p^2+ {\tilde M}^2 }\equiv \frac{E}{2},
\label{INCMOM1}
\earr
where $E$ is the center of mass energy.
Since we have momentum conservation only in the perpendicular direction we expect that the projection on the interface of the  direction of the incident center of mass will  rotate by an angle $\phi$. Thus for the outgoing particles we parametrize  the momenta as follows
\begin{equation}
\mathbf{p'}_1=(p'\cos \phi ,\;p'\sin \phi ,\;p'_{1}), \qquad 
\mathbf{p'}%
_{2}=(-p'\cos \phi ,\;-p'\sin \phi ,\;p'_{2}),\qquad
|\mathbf{p'}_{1\perp }|=p',
\label{OUTMOM}
\end{equation}%
which satisfies $\mathbf{p'}_{1\perp }=-\;%
\mathbf{p'}_{2\perp }$ and $|\mathbf{p'}_{1\perp }|= |\mathbf{p'}_{2\perp }|=p'$. Let us emphasize that $p_{1,2}$ denote the third component of the momenta $\mathbf{p}_{1,2}$ and should not be confused with $|\mathbf{p}_{1,2}|$, with analogous convention for the primed vectors.  As we will show in the next section, the exchange of the $\theta$-particle will not provide the momentum conserving $\delta$ function in the $z$-direction $\delta(p_1'+p_2')$ so that $p_1'$ and $p_2'$ remain arbitrary within the restriction imposed by energy conservation. Our aim is to determine the probability of occurrence for each combination $(p_1', \, p_2')$, which will be provided by the transition amplitude. 

The energies are
\beq
E'_1= \sqrt{p'^2 +p'_{1}{}^2+{\tilde M}^2}, \qquad  E'_2= \sqrt{p'^2 +
p'_{2}{}^2 +{\tilde M}^2}.
\label{EOUT}
\eeq
Energy conservation $E_1+E_2=E=E'_1+E'_2$ allows to solve $p'={\cal P}'$ in terms of $E, p'_1, p'_2$ with the result
\begin{equation}
{\cal P'}^{2}=\frac{E^{ 2}}{4}\left[ \left( 1-\left( \frac{p'_{2}{}^2}{%
E^{ 2}}+\frac{p'_{1}{}^2}{E^{ 2}}\right) \right) ^{2}-\allowbreak
4\left( \frac{{\tilde M}^{2}}{E^{ 2}}+\frac{p'_{2}{}^2}{E^{ 2}}\frac{%
p'_{1}{}^2}{E^{ 2}}\right) \right].
\label{PP2}
\end{equation}%
Since ${\cal P '}^{2} \geq 0$ and the right hand side of (\ref{PP2}) is not manifestly positive definite, this relation constraints the region where the momentum violation can occur. In other words, the allowed $p'_1$ and $p'_2$  must be such that ${\cal P '}^{2} \geq 0$.
 
The outgoing energies can be  written in a convenient form as 
\begin{equation}
E'_{2}=\frac{\left( E^2+p'_{2}{}^2-p'_{1}{}^2\right) }{2E }, \qquad 
E'_{1}=\frac{\left( E^2+p'_{1}{}^2-p'_{2}{}^2\right) }{2E },
\label{CONVOUTE}
\end{equation}
and it is a direct matter to verify $E'_{1}+E'_{2} =E$.

In preparation for the definition of a modified cross section we observe that we have  three delta functions that can be integrated, which  enforce the conservation laws at our disposal: energy and perpendicular momentum conservation.  This means  that from the six original outgoing variables ${\mbf p}_1'$ and $ {\mbf p}_2'$ we will be left with three undetermined parameters, in contrast with the two variables left when we have full energy-momentum conservation. We choose them to be $p'_1, p'_2$ and the angle $\phi$ measuring the rotation of the projection of the center of mass direction on the interface.  The required integral in the outgoing phase space will be of the form 
\begin{equation}
dI=d^{3}{\mbf p'}_1 \, d^{3}{\mbf p'}_2\;\delta ^{2}(\mathbf{p'}_{1\perp }+\mathbf{p'}_{2\perp
})\delta (E'_{1}+E'_{2}-E)\,  {\cal F},
\label{DEFF}
\end{equation}
where the function ${\cal F}$, depending on all the variables, need not to be  specified for the moment. This can  naturally be split into
\begin{equation}
dI=\left( d^2{\mbf p'}_{1 \perp }dp'_{1}\right) \left( d^2
{\mbf p'}_{2\perp }dp'_{2}\right)
\;\delta ^{2}(\mathbf{p'}_{1\perp }+\mathbf{p'}_{2\perp })\delta
(E'_{1}+E'_{2}-E)\,\, {\cal F}.
\label{SPLIDEFF}
\end{equation}%
Integrating  $d^{2}{\mbf p'}_{1\perp }$ yields
\begin{equation}
dI=dp'_{1} \;dp'_{2}\, d^{2} {\mbf p'}_{2\perp }\;\delta (E'_{1}+E'_{2}-E)\,{\cal F} \qquad \mathbf{p'}%
_{1\perp }=-\;\mathbf{p'}_{2\perp }\quad \mathbf{p'}_{1\perp }^{2}=%
\mathbf{p'}_{2\perp }^{2}=p'^{2}. 
\label{INTP1}
\end{equation}
Next we set $ d^2{\mbf p'}_{2\perp }=p'\;dp'\;d\phi$ and rewrite 
\begin{equation}
dI=dp'_{2}\, dp'_{1}\; p'\; dp'\; d\phi \;
\delta \left(\sqrt{p'^{2}+ p'_{1}{}^2+{\tilde M}^{2}}+\sqrt{%
p'^{2}+p'_{2}{}^2+{\tilde M}^{2}}-E \right) \ {\cal F}.
\label{SPLITP2}
\eeq
We carry out the integral with respect to $p'$ and we use \begin{equation}
\delta \left(\sqrt{p'^{2}+ p'_{1}{}^2+{\tilde M}^{2}}+\sqrt{%
p'^{2}+p'_{2}{}^2+{\tilde M}^{2}}-E \right) =\frac{E'_{1}E'_{2}}{p' E}\delta (p'-{\cal P}')
\label{INTPP}
\end{equation}%
where ${\cal P}'$ was already calculated in Eq. (\ref{PP2}).
The final result can be presented as 
\beq
\frac{d^3 I}{dp'_1 dp'_2 d \phi}=\frac{E'_1 \, E'_2}{E} {\cal F},
\label{DIFMCS1}
\eeq
with $E'_1$ and $E'_2$  given by Eq. (\ref{CONVOUTE}). The remaining functions in  ${\cal F}$ are evaluated in terms of the independent variables  $p'_1, p'_2, \phi$ according to the relations previously developed. Notice that (\ref{DIFMCS1}) provides a distribution of the momentum violating events  precisely in terms of the undetermined components in the direction $z$ of the outgoing momenta, together with the rotation angle  of the projection of the center of mass direction on the interface.

\subsection{ The transition (scattering) amplitude}

\label{IV2}
The dynamics of the process is given by the transition amplitude 
to second order in  $\tilde{\lambda}$
\begin{equation}
\langle f|(i\, \mfT)| i \rangle={\frac{1}{2}}\langle{\mbf{p}'_1,\mbf{p}'_2}|\mathcal{T}\left\{
(-i{\tilde \lambda})\Big(\int d^4x \chi^{*}\chi \Phi\Big)(-i{\tilde \lambda})\Big(\int d^4x' \chi^{*}\chi\Phi\Big) \right\}|{\mbf{p}_1,\mbf{p}_2}\rangle,
\label{86}
\end{equation}
which is calculated in detail in the Supplementary Materials \cite{SuppMaterials}. As  shown in Eq. (\ref{propagador2}), the propagator mediating the interaction can be separated into
\beq
\Delta_F=\Delta_{F 0} (x-x')+ \Delta_{F  {\tilde \theta}}(x,x'),
\label{SPLITFEYN} 
\eeq
where the first term is the usual contribution that conserves momentum, while the second expression contains the momentum-violating part.
We will focus only in the momentum-violating sector, such that $\delta(p_{1}+p_{2}-p'_{1}-p'_{2})=0$, which it is encoded in the amplitude 
\beq
\langle f|i\rangle^{NC}=\langle f|(i \, \mfT_{\tilde \theta})|i\rangle=(2\pi) \delta^0(P^0_f-P^0_i) (2\pi)^2 \delta^2({\mbf P}_{f \perp}-{\mbf P}_{i \perp})(i {\cal M}^{NC})\equiv {\mft}_{\tilde \theta},
\label{NEWAMP}
\eeq
which now is lacking the delta function enforcing momentum conservation in the $z$ direction. From Eq. (\ref{86}) and inserting $\Delta_{F  {\tilde \theta}}(x,x')$ we obtain  
\begin{eqnarray}
{\mft}_{{\tilde  \theta}}&=& {\tilde \lambda}^2\int d^4xd^4x'\frac{\tilde{\theta}}{2}
\int_{-\infty}^{+\infty} \frac{d^4k}{(2\pi)^4}\frac{e^{-ik^0(x^0-x'^0)+i\mbf{k}_{\perp}\cdot(\mbf{x}-\mbf{x'})_{\perp}}}{(k^2-m^2+i\epsilon)(k-i\frac{\tilde{\theta}}{2})}e^{ik(|z|+|z'|)}\nonumber\\
&&\quad \times e^{ip'_2\cdot x}e^{-ip_2\cdot x}e^{ip'_1\cdot x'}e^{-ip_1\cdot x'}+\left( \mbf{p}'_1\leftrightarrow\mbf{p}'_2\right),
\label{THETAT}
\end{eqnarray}
in the notation of Eq. (\ref{NOTATION}).
Let us introduce
\beq
q_1^\mu=p'_1{}^\mu-p_1{}^\mu, \qquad q_2^\mu=p'_2{}^\mu-p_2{}^\mu, \qquad {\tilde q}_1^\mu=p'_2{}^\mu-p_1{}^\mu, \qquad {\tilde q}_2^\mu=p'_1{}^\mu-p_2{}^\mu,
\label{QQTILDE}
\eeq 
where  ${\tilde q}_1^\mu$ and ${\tilde q}_2^\mu$ denote the momentum transfer in the crossed channel.  Their explicit expressions are
\barr
&&q_1^\mu =p'_{1}{}^\mu-p_{1}{}^\mu=(E'_1-E_1, \, p'\cos \phi-p\sin \theta, \, p'\sin\phi, \, p'_1-p\cos\theta). \notag \\
&&q_2^\mu =p'_{2}{}^\mu-p_{2}{}^\mu=(E'_2-E_2, \, -p'\cos \phi+p\sin \theta, \, -p'\sin\phi, \, p'_2+p\cos\theta). 
\label{q} \\
&&{\tilde q}_1^\mu =p'_{2}{}^\mu-p_{1}{}^\mu=(E'_2-E_1, \,- p'\cos \phi-p\sin \theta, \, -p'\sin\phi, \, p'_2-p\cos\theta). \notag \\
&&{\tilde q}_2^\mu =p'_{1}{}^\mu-p_{2}{}^\mu=(E'_1-E_2, \, p'\cos \phi+p\sin \theta, \, p'\sin\phi, \, p'_1+p\cos\theta). 
\label{crossq}
\earr
 In general we denote the crossing operation over momentum dependent quantities with a tilde over the quantity to which it is applied.

Following the steps indicated in the Supplementary Materials \cite{SuppMaterials}  we arrive at 
\barr
&&{\mft}_{\tilde \theta} =i \lambda^2 \tilde{\theta} \frac{Q}{(Q-i\frac{\tilde{\theta}}{2})}(2\pi)^3\delta(P_f^0-P_i^0)\delta^2(\mbf{P}_{f \perp}-\mbf{P}_{i \perp})\frac{1}{\left[(q_2)^2-Q^2+i\epsilon\right]\left[(q_1)^2-Q^2+i\epsilon\right]} \notag  \\
&&\hspace{2.4cm} +\left( \mbf{p}'_1\leftrightarrow\mbf{p}'_2\right),
\label{T_theta6}
\earr
with 
\beq
Q\equiv\sqrt{ (q_1^0)^2-(\mbf{q}_{1\perp})^2-m^2}=\sqrt{(q_2^0)^2-(\mbf{q}_{2,\perp})^2-m^2}.
\eeq

Again we remark  that $q_1$ an $q_2$ in Eq. (\ref{T_theta6}) denote the $z$ component of the corresponding four-momenta $q_1^\mu$ and $q^\mu_2$, respectively. In the following  we denote $q_\mu q^\mu\equiv q\cdot q$ to avoid confusion with the standard $q^2$ notation. 

From Eqs. (\ref{NEWAMP}) and  (\ref{T_theta6}) we read
\begin{equation}
\begin{split}
{\cal M}^{NC} =& {\tilde \lambda}^2 {\tilde{\theta}} \frac{Q}{(Q-i\frac{\tilde{\theta}}{2})}\, \frac{1}{\left[q_2 \cdot q_2-m^2+i\epsilon\right]\left[q_1 \cdot q_1-m^2+i\epsilon\right]}+ \left( \mbf{p}'_1\leftrightarrow\mbf{p}'_2\right),
\end{split}
\label{T_theta1}
\end{equation}
which yields 
the total transition probability per unit time
\beq
 d \dot{P}=\frac{(2\pi )^{3}\delta^0(P^0_f-P^0_i) \delta^2({\mbf P}_{f \perp}-{\mbf P}_{i \perp})\;}{4E_{1}^{\prime }E_{2}^{\prime
}\;\;4\;E_{1}\;E_{2}}  \left|{\cal M}^{NC}\right|^{2}\frac{1}{L^{4}}\frac{d^{3}{\mbf p}'_{1}}{\left( 2\pi \right)
^{3}}\frac{d^{3}{\mbf p}'_{2}}{\left( 2\pi \right) ^{3}},  
\label{TOTPROB11}
\eeq
as shown in the Supplementary Materials \cite{SuppMaterials}. 
To impose periodic boundary conditions we take a cubic box $L^3$ such that the interface $z=0$ is parallel to its  two opposite  far away faces, and  where we have one incident particle  with density $\rho=1/L^3$. We take $L$ much larger that the De Broglie wave length of the incident particles, i.e. $L \geq h/|{\mbf p}_1|$.   Also, we  defined our coordinate system such that direction of the incident particles moving collinearly  forms an angle $\theta$ with the normal to the interface. Notice that the number of particles per unit time impinging per unit area perpendicular to the relative direction of motion is always $\rho v_{\rm rel}$,  independently of the angle of incidence. Also let us  observe that even in the case $\theta=\pi/2$, i.e. when $p_1=p_2=0$, momentum non-conservation  can dynamically  yield ${\cal M}^{NC} \neq 0$ thus producing a non-zero transition probability to some $(p'_1, p'_2)$.   
To define the  analogous to the standard cross section, which is independent of the box regularization length $L$,  we need to specify an alternative definition of the incoming flux. This is because  the lack of a $\delta(P_{fz}-P_{if})$ in (\ref{NEWAMP}) modifies the dimensions
of ${\cal M}^{NC}$. To this end we choose to  define 
a flux per unit length 
\beq
{\tilde F}=\frac{\rho \,  v_{\rm rel} }{L}=\frac{v_{\rm rel}}{L^4}.
\label{NEWFLUX}
\eeq
To measure $\tilde F$ we consider an area $A$ perpendicular to the direction of the incident ${\mathbf v}_{\rm rel}$, count the particles per unit time ${\dot N}$ crossing the area and divide by $A^{3/2}$, obtaining ${\tilde F}={\dot N}/A^{3/2}$.
Using  $\tilde F$  we can define  
an  observable measure of the momentum violating  transition probability as
\beq
d \, \Xi \equiv  \, \frac{d {\dot P}}{{\tilde F}}= d {\dot P}\,\frac{1}{v_{\rm rel}} L^4,
\label{TRANSPROB}
\eeq
having the dimensions of ${\rm cm}^3$ and being independent of $L$.
Substituting in (\ref{TOTPROB1}) we obtain
\beq
 d \, \Xi=\frac{(2\pi )^{3}\delta^0(P^0_f-P^0_i) \delta^2({\mbf P}_{f \perp}-{\mbf P}_{i \perp})\;}{4E_{1}^{\prime }E_{2}^{\prime
}\;\;4\;E_{1}\;E_{2}}  \left|{\cal M}^{NC}\right|^{2}\frac{1}{v_{\rm rel}}\frac{d^{3}{\mbf p}'_{1}}{\left( 2\pi \right)
^{3}}\frac{d^{3}{\mbf p}'_{2}}{\left( 2\pi \right) ^{3}} 
\label{TOTPROB1}
\eeq
We can perform the integration over the delta functions following the previous steps  yielding Eq. (\ref{DIFMCS1}). From Eq. (\ref{DEFF}) we identify ${\cal F}$ as
\beq
{\cal F}=\frac{1}{128 \pi^3}\frac{1}{E_{1}^{\prime }E_{2}^{\prime
}\;\;E_{1}\;E_{2}}  \frac{1}{v_{\rm rel}}\, \left|{\cal M}^{NC}\right|^{2}
\label{DEFF1}
\eeq
and obtain the final  result
\beq
\frac{d^3 \Xi}{dp'_1 dp'_2 d \phi}=\frac{1}{32 \pi^3}\frac{1}{|{\mathbf p}|E^2} \, \left|{\cal M}^{NC}\right|^{2}
\label{DIFMCS}
\eeq
in the ingoing center of mass, where  $E_1=E_2= \sqrt{{\mbf p}^2+ {\tilde M}^2}= E/2$ and $v_{\rm rel}= 2|{\mathbf p}|/E$.
The dependence on the remaining variables is contained in $|{\cal M}^{{ NC}}|^2 \, $ written in Eq. (\ref{T_theta1}), where now $p'={\cal P}'$ since we have performed the integration in (\ref{INTPP}). Their explicit expression in terms of the known initial quantities $p,\, E, \, \theta, \,{ \tilde M} $, together with the outgoing variables
 $p'_1$, $p'_2$ and $\phi$ are 
\begin{eqnarray}
q_1 \cdot q_1&=&\left(\frac{p'_{1}{}^2-p'_{2}{}^2}{2E} \right)^2-\left({{\cal P}'}^2-2{{\cal P}'}p\cos\phi\sin\theta+p^2\sin^2\theta \right)-(p'_{1}-p\cos\theta)^2, \label{105}\\
q_2 \cdot q_2&=&\left(\frac{p'_{1}{}^2-p'_{2}{}^2}{2E} \right)^2-\left({{\cal P}'}^2-2{{\cal P}'}p\cos\phi\sin\theta+p^2\sin^2\theta \right)-(p'_{2}+ p\cos\theta)^2,\\
 {\tilde q}_1 \cdot {\tilde q}_1&=&\left(\frac{p'_{1}{}^2-p'_{2}{}^2}{2E} \right)^2-\left({{\cal P}'}^2+ 2{{\cal P}'}p\cos\phi\sin\theta+p^2\sin^2\theta \right)-(p'_{2}-p\cos\theta)^2,\\
{\tilde q}_2 \cdot {\tilde q}_2&=&\left(\frac{p'_{1}{}^2-p'_{2}{}^2}{2E} \right)^2-\left({{\cal P}'}^2+ 2{{\cal P}'}p\cos\phi\sin\theta+p^2\sin^2\theta \right)-(p'_{1}+p\cos\theta)^2, \label{106}\\
Q^2&=&\left(\frac{p'_{1}{}^2-p'_{2}{}^2}{2E} \right)^2-\left({{\cal P}'}^2-2{{\cal P}'}p\cos\phi\sin\theta+p^2\sin^2\theta \right)-m^2, \\
{\tilde Q}^2&=& \left(\frac{p'_{1}{}^2-p'_{2}{}^2}{2E} \right)^2-\left({{\cal P}'}^2+ 2{{\cal P}'}p\cos\phi\sin\theta+p^2\sin^2\theta \right)-m^2, \\
\mathcal{P}'&=&\frac{1}{2E}\sqrt{\left(E^2-(p'_{1}{}^2+p'_{2}{}^2)\right)^2-4\left(E^2 {\tilde M}^2+p'_{1}{}^2\, p'_{2}{}^2 \right)}.\label{111}
\end{eqnarray}

\subsection{Regions where momentum is not conserved }

\label{IV3}

In this section we determine  the allowed region in the $p_1'$-$p_2'$ plane where the violation of the $z$-component of the momentum occurs.
Also we locate the loci where the scattering amplitude diverge identifying  the zeros in the denominators of  Eq. (\ref{T_theta1}).

The condition ${{\cal P}'}^2>0$  determines the existence of momentum-violating events in  the ($p'_1$-$p'_2$)  plane. To characterize the acceptable  region we introduce a convenient parametrization defined by
\beq
\frac{p'_1}{E}= \rho \cos \alpha \equiv X, \qquad \frac{p'_2}{E}=\rho \sin \alpha \equiv Y, \qquad {\mathfrak b}= \frac{p}{E}, \qquad  {\mathfrak c}= \frac{{\tilde M}}{E}, \qquad  {\mathfrak b}^2+ {\mathfrak c}^2= \frac{1}{4}. 
\label{DEFPARAM} 
\eeq
As shown in the Supplementary Materials \cite{SuppMaterials}, energy conservation provides the  necessary condition $\rho <1 $ for the allowed region. The sufficient condition  comes  from Eq. (\ref{PP2}). Using the  quantities defined in (\ref{DEFPARAM}) we can write ${{\cal P}'}^2 \geq 0$ as
\beq
W(\rho, \alpha)\equiv 4 \mfb^2 -2\rho^2 + \rho^4 \cos^2 2 \alpha  \geq 0.
\label{CONDPOS}
\eeq
For a given $\alpha$, the value of $\rho$ establishing  the limit between the allowed and the forbidden region is given by the solutions of $W(\rho, \alpha)=0$, which are 
\beq
\rho^2_+=\left|\frac{1}{\cos^2 2\alpha}\left(1 + \sqrt{1-4 \mfb^2\cos^2 2\alpha}  \right)   \right |, \quad 
\rho^2_-=\left|\frac{1}{\cos^2 2\alpha}\left(1 - \sqrt{1-4 \mfb^2\cos^2 2\alpha}  \right) \right |.
\label{RHOPM}
\eeq
We find that  the  relation $\rho_- < 1 < \rho_+$ is satisfied, so that the allowed region is determined by $\rho_-$.  To decide whether this  region is either $\rho>\rho_-$ or $\rho < \rho_-$  we consider the behavior of the function $W(\rho, \alpha)$ in Eq. (\ref{CONDPOS}), which must be positive and satisfies $W(\rho_-, \alpha)=0$. The result $\frac{\partial W(\rho, \alpha)}{\partial \rho^2}=-2(1-\rho^2 \cos^2 2\alpha)<0$ shows that $W(\rho, \alpha)$ is a decreasing function of $\rho$ in the range $ \rho <1$. That is to say, the maximum value $W=4{\mathfrak b}^2 >0 $ 
occurs at $\rho=0$ and the minimum is $W=0$ at $\rho = \rho_-$. 
Thus the allowed region for momentum violation at a given angle 
$\alpha $ is for $ \rho < \rho_-(\alpha)$.   Let us remark that in the limit $\cos 2 \alpha\rightarrow 0$ we have $\rho_-\rightarrow  \sqrt{2} \, \mfb$.

Next  we focus on the poles of  the scattering amplitude ${\cal M}^{NC}$ to determine whether or not this amplitude blows up inside the allowed momentum region. Nevertheless, given the complexity of these denominators resulting from  Eqs. (\ref{105}-\ref{106}), we further impose the additional restriction  $\theta=0$ in order to simplify our results. That is to say, we only consider 
the case when the incoming particles impinge along the $z$ axis, perpendicularly to the interface. Then, the Eqs. (\ref{105}-\ref{106}) yield
\begin{eqnarray}
&&
\frac{4\left( q_{1}\cdot q_{1}-m^{2}\right)}{E^{2}} = 2\rho^2 -1-4\left( \rho \cos \alpha-\mathfrak{b} \right) ^{2}+ \Delta \equiv D_1 , \nonumber \\ 
&&
\frac{4\left( q_{2}\cdot q_{2}-m^{2}\right)}{E^{2}} = 2\rho^2 -1-4\left( \rho \sin \alpha+\mathfrak{b}\right)^{2}+ \Delta \equiv D_2,  \nonumber \\
&& \frac{4\left( \tilde{q}_{1}\cdot \tilde{q}_{1}-m^{2}\right)}{E^2} =2\rho^2 -1
-4\left( \rho \sin \alpha-\mathfrak{b} \right) ^{2} +  \Delta \equiv D_3,  \nonumber \\ 
&&
\frac{4\left( \tilde{q}_{2}\cdot \tilde{q}_{2}-m^{2}\right)}{E^{2}} =2\rho^2 -1-4 \left( \rho \cos \alpha 
+\mathfrak{b} \right) ^{2} +  \Delta \equiv D_4, 
\label{DENPOLAR1}
\end{eqnarray}
in terms of the polar variables. Here  
\beq
\Delta \equiv \frac{4({\tilde{M}}^{2}-m^{2})}{E^{2}}=4\mfc^2-4\mfr^2, \qquad \mfr=\frac{m}{E} .
\label{PARAMFIN}
\eeq

Now we are interested in finding the loci where the scattering amplitude diverges, which will be presented in a series of polar plots in the following. This  requires to solve for $\rho$ at a given angle in each of the four  equations  $D_i=0, \,\, i=1,2,3,4$ in (\ref{DENPOLAR1}). We warn the reader that these results   may exhibit both negative  as well as imaginary values,  which must be adequately incorporated  in these plots which only require the absolute value of $\rho$ in the regions where the solutions are real. 
Fortunately, the number of equations to be considered  can be restricted due to additional symmetries .  From Eqs. (\ref{DENPOLAR1}) in Cartesian form we verify
\beq
D_1(-X)=D_4(X), \qquad D_2(-Y)=D_3(Y).
\label{SYMM1}
\eeq
 In other words, it is enough to deal with only two equations  arising either from $(D_1, D_2)$, or from $(D_3, D_4)$, thus reducing the search of the original eight values of $\rho$ to only four. 
Using the the polar parametrization (\ref{DEFPARAM}), together with (\ref{PARAMFIN}), 
we choose to deal with 
\barr
&&\rho_1=\frac{1}{2 \cos 2\alpha}\left(4 \mfb \cos \alpha +\sqrt{16 \mfb^2 \sin^2 \alpha-8 \mfr^2 \cos  2 \alpha }\right), \nonumber \\
&&\rho_2=\frac{1}{2 \cos 2\alpha}\left(4 \mfb \cos \alpha -\sqrt{16 \mfb^2 \sin^2 \alpha-8 \mfr^2 \cos  2 \alpha }\right), \nonumber \\
&&\rho_3=\frac{1}{2 \cos 2\alpha}\left(4 \mfb \sin \alpha -\sqrt{16 \mfb^2 \cos^2 \alpha-8 \mfr^2 \cos  2 \alpha }\right), \nonumber \\
&&\rho_4=\frac{1}{2 \cos 2\alpha}\left(4 \mfb \sin \alpha +\sqrt{16 \mfb^2 \cos^2 \alpha-8 \mfr^2 \cos  2 \alpha }\right).
\label{CEROSDEN}
\earr
As a matter of fact, $\rho_1$ and $\rho_2$ are the two solutions of $D_1=0$, while $\rho_3$ and $\rho_4$ arise from $D_2=0$.  

An additional symmetry is present due to the shift $\alpha \rightarrow \alpha +\pi$ in (\ref{CEROSDEN}), which  yields 
\barr
 && \rho_1(\alpha+\pi)=-\rho_2(\alpha), \qquad
 \rho_3(\alpha+\pi)=-\rho_4(\alpha). 
 \label{ADDSYMM}
\earr
This indicates that it is enough to consider only a pair of solutions: either $(\rho_1, \rho_3)$ or
$(\rho_2, \rho_4)$ and subsequently rotate the plot by $\pi$ to obtain the complete description. Next we show some particular examples. 
\subsubsection{The case $\mfr=0$}
The simplest case is to consider 
arbitrary values for  $\mfb=p/E$, while setting $m=0, (\mfr=0)$. Then, the required values of $\rho_1$ and $\rho_3$ from Eqs. (\ref{CEROSDEN}) reduce to
\barr
&&\rho_1(\alpha)=\frac{2 \mfb}{\cos 2 \alpha}\left( \cos\alpha +|\sin \alpha|\right), \label{RHO10} \\
&&\rho_3(\alpha)=\frac{2 \mfb}{\cos 2 \alpha}\left( \sin \alpha -
|\cos \alpha|\right),
\label{RHO30}
\earr
where $\sqrt{\sin^2 \alpha}=|\sin \alpha|$,  and similarly for $\sqrt{\cos^2 \alpha}$. As we can see,   no imaginary parts appear in the solutions $\rho_i(\alpha)$. 
\begin{figure}[h!]
\centering 
\includegraphics[scale=0.5]{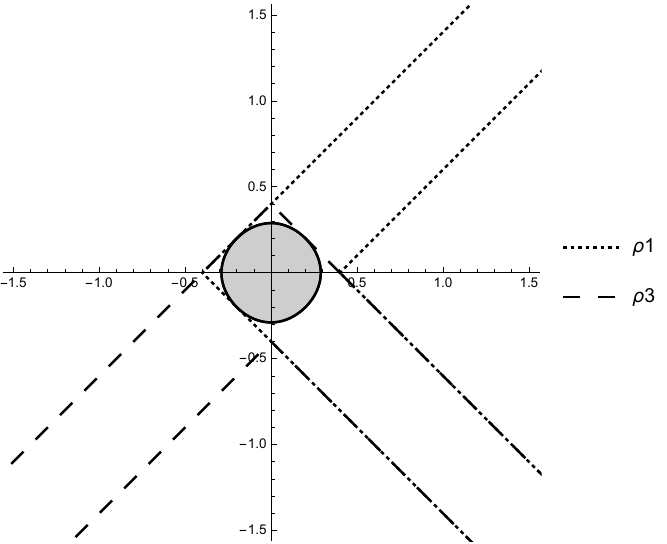} \quad \qquad \qquad 
\includegraphics[scale=0.5]{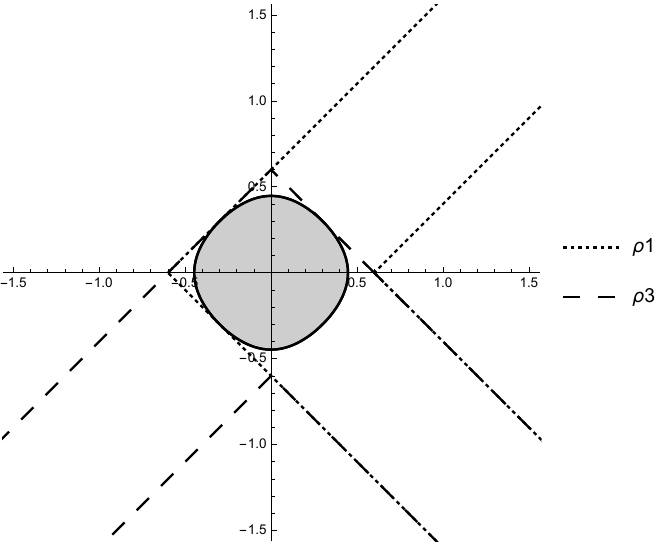}
\caption{Polar plots for $|\rho_1|$  (dotted line) and $|\rho_3|$ (dashed line), for the choices 
$\mfb=0.2, \, 0.3$, from left to right. The shaded area is the allowed region where momentum violation in the third direction occurs.}
\label{FIG0}
\end{figure}
\begin{figure}[h!]
\includegraphics[scale=0.5]{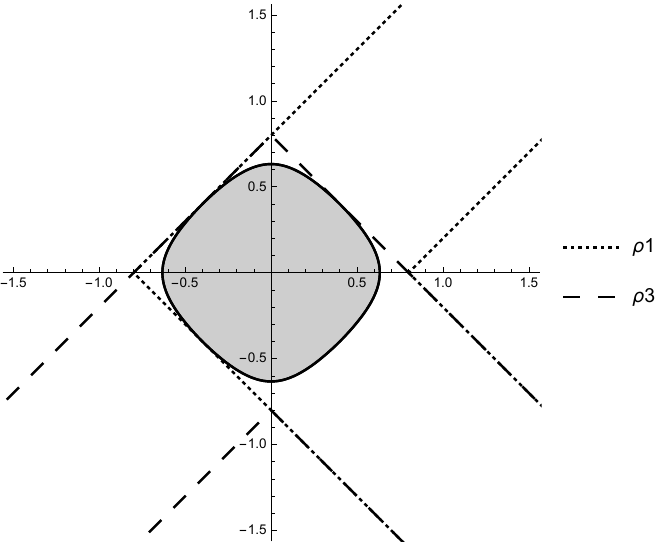} \quad \qquad  \qquad %
\includegraphics[scale=0.5]{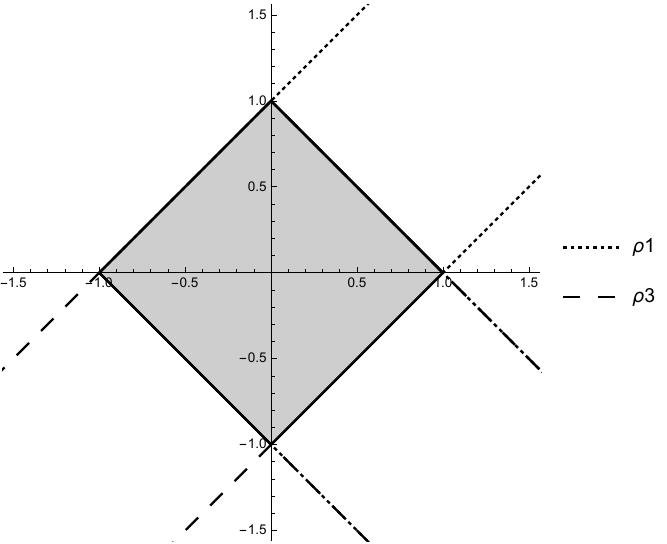}
\caption{Polar plots for $|\rho_1|$ (dotted line) and $|\rho_3|$ (dashed line), for the choices 
$\mfb= 0.4, \, 0.5 $, from left to right. The shaded area is the allowed region where momentum violation in the third direction occurs.}
\label{FIG1}
\end{figure}

Polar plot of the loci  $|\rho_1|$ and $|\rho_3|$, including the boundary of the  allowed region $\rho < \rho_-$, for the values  $\mfb=0.2, \, 0.3, \, 0.4, \, 0.5 $  are shown in Figs. \ref{FIG0} and \ref{FIG1}. We have  $|\rho_1(0)|=|\rho_3(0)|=2 \mfb$. We observe that the boundary of the allowed region is always tangent to the loci of the poles of the scattering amplitude at the angles $\alpha_n=(2n+1)\pi/4, \, \,  n=0,1,2\, $, where $\cos 2 \alpha_n=0$. This is verified by  taking the limit $\alpha \rightarrow \alpha_n$ in the second expression (\ref{RHOPM}) for $\rho_-$, where we find $\rho_-(\alpha_n)=\sqrt{2} \, \mfb$. As an additional check, from Eqs. (\ref{RHO10}-\ref{RHO30}) we realize that these touching points at $\alpha= \pi/4,\, 3\pi/4, \, 5\pi/4 $, yield $\rho=\sqrt{2} \mfb   $. Recall the full plot is obtained by adding to the displayed figures their  rigid rotation by the angle $\pi$. 

\

\subsubsection{The case of arbitrary $\mfr$} 

Next we  investigate the location of poles of ${\cal M}^{NC}$  when considering   $0< \mfr < \infty$. This case is more subtle because the values of $\rho$ may acquire  imaginary parts from the square roots in Eqs. (\ref{CEROSDEN}). The corresponding angular region  must be excluded from the solution yielding   the sudden interruption of the plotted lines in some graphs. The radius $\rho$ are finite at these points. For clarity in the figures, here we choose to plot separately $\rho_1$  and $\rho_3$. 

\begin{figure}[h!]
\centering 
\includegraphics[scale=0.5]{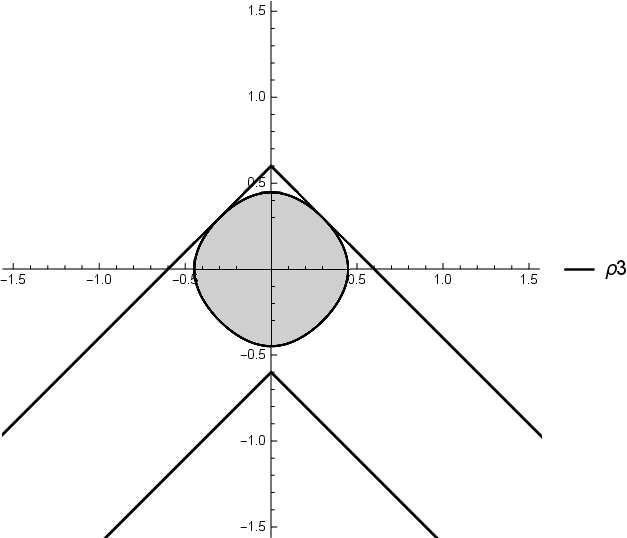} \hspace{2cm}
\includegraphics[scale=0.5]{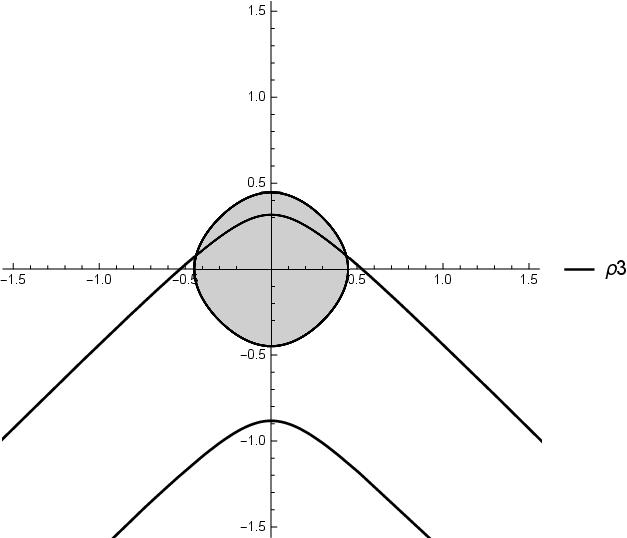} %
\caption{Polar  plots for $|\rho_3|$ at fixed $\mfb=0.3$ with  $\mfr=0$ and $\mfr=0.2$, from left to right.}
\label{RHO33}
\end{figure}
\begin{figure}[h!]
\includegraphics[scale=0.5]{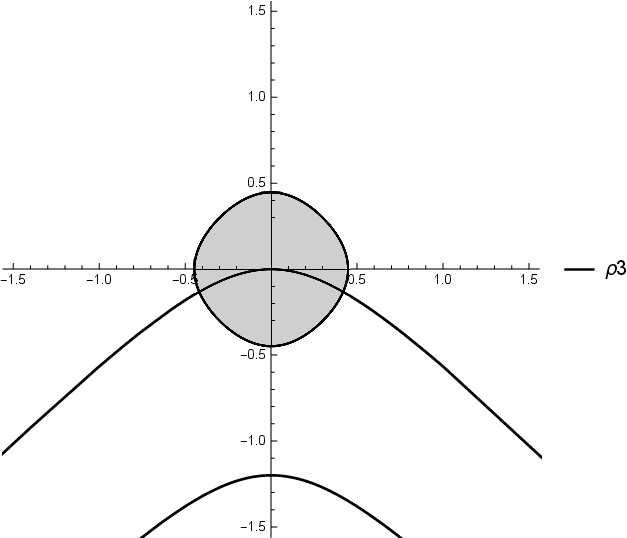} \hspace{2cm}
\includegraphics[scale=0.5]{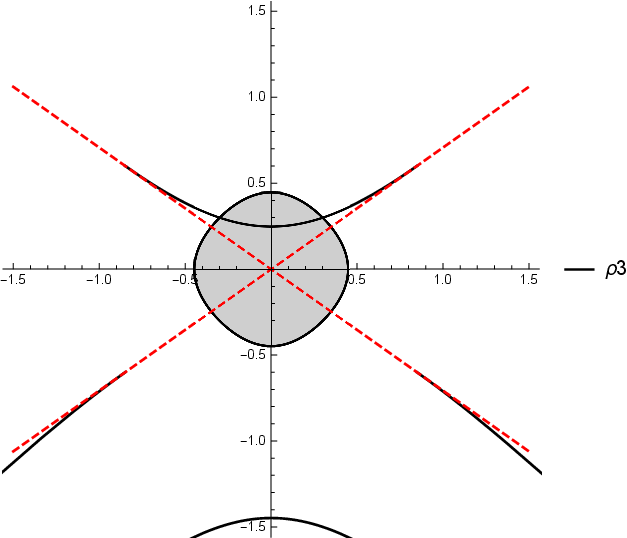}
\caption{Polar  plots for $|\rho_3|$ at fixed $\mfb=0.3$ with
  $\mfr=\mfr_0=0.424$ and $\mfr=0.6 $, from left to right.}
\label{RHO3}
\end{figure} 

Looking at  $\rho_3$ we find the critical value $\mfr_0=\sqrt{2}\,\mfb$ below which $\rho_3$  is always real. An imaginary part arises  when  $\mfr > \mfr_0$ and falls within   the angular 
ranges $0 < \alpha <\alpha_{c3}, \, \pi-
\alpha_{c3}<\alpha < \pi+\alpha_{c3}, \,  2\pi-\alpha_{c3} < \alpha < 2 \pi  $,  with 
\beq
\alpha_{c3}=\arccos\left[\sqrt\frac{\mfr^2}{2(\mfr^2-\mfb^2)}\right].
\label{ANGR1}
\eeq
This defines a forbidden angular region. Let us observe that for a given value of $\mathfrak{b}$, we have  $0 < \alpha_{c3} < \pi/4$ in the interval $\sqrt{2}\mathfrak{b}<\mathfrak{r}< \infty$.

We illustrate this case in Figs. \ref{RHO33} and \ref{RHO3}, for $\mfb=0.3$, when $\mfr_0=0.424$. The polar plots are for $\mfr=0, \, 0.2, \, \mfr=\mfr_0$ and $\mfr=0.6$.  The limits of the  forbidden angular region are indicated by the dashed lines in the  plot to the right of  Fig. \ref{RHO3}. We verify  that as increasing $\mfr$, the plot for $|\rho_3|$ moves  down  until we get to $\mfr_0$ when the upper  line of the graph just touches the $x$-axis. Further increase in $\mfr$ produces imaginary parts and the allowed loci start receding from the origin falling outside the boundaries of the forbidden angular region determined by $\alpha_{c3}$. On the other hand, the increase in $\mfr$ makes the upper branch describing the poles to enter into the allowed momentum non-conservation region indicated by the dashed area.

In the case of $\rho_1$, for a given $\mfb$, we  have an imaginary part for arbitrary values of $\mfr > 0$ in the angular ranges $0 < \alpha <\alpha_{c1}, \, \pi-
\alpha_{c1}<\alpha < \pi+\alpha_{c1}, \,  2\pi-\alpha_{c1} < \alpha < 2 \pi  $,  with
\beq
\alpha_{c1}=\arcsin\left[\sqrt\frac{\mfr^2}{2(\mfr^2+\mfb^2)}\right].
\label{ANGR2}
\eeq 
Similarly to the previous case, we  have  $0 < \alpha_{c1} < \pi/4$ in the interval $0 < \mathfrak{r}< \infty$, for a fixed  value of $\mathfrak{b}$.

The limits of the new forbidden angular  region are indicated by the dashed lines in the Figs. \ref{RHO11} and \ref{RHO1}. We observe  that when increasing $\mfr$ the loci of the poles of the scattering amplitude move  away from the center and do not penetrate the allowed momentum-violating region ( dashed area). The polar plots in Figs. \ref{RHO11} and \ref{RHO1} are for fixed $\mfb=0.3$ with  $\mfr=0, \, \mfr=0.2, \,\mfr=\mfr_0$ and $ \mfr=0.6 $.

\begin{figure}[h!]
\centering 
\includegraphics[scale=0.5]{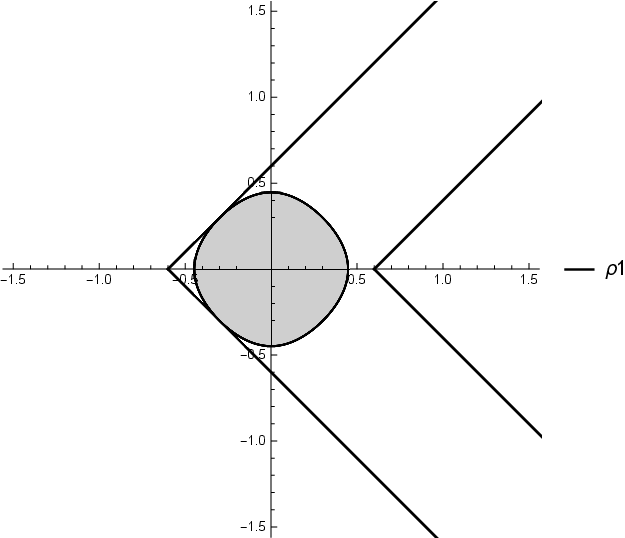} \hspace{.2cm}
\includegraphics[scale=0.5]{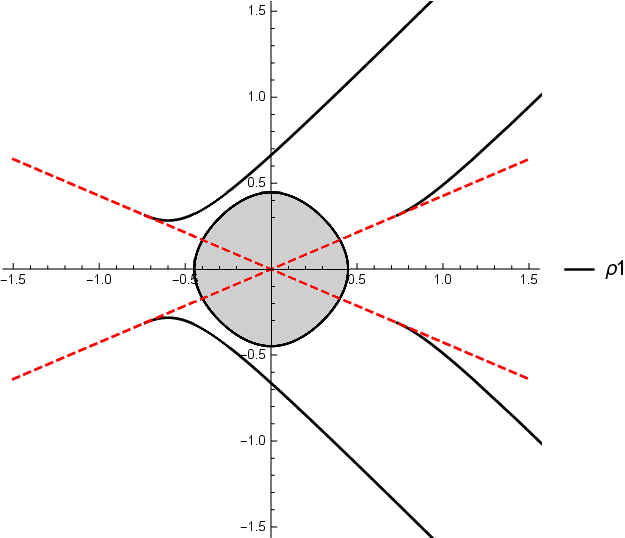}  %
\caption{Polar  plots for $|\rho_1|$ at fixed $\mfb=0.3$ with  $\mfr=0, \,\mfr=\mfr=0.2 $, from left to right.} \label{RHO11}
\end{figure}
\begin{figure}[h!]
\includegraphics[scale=0.5]{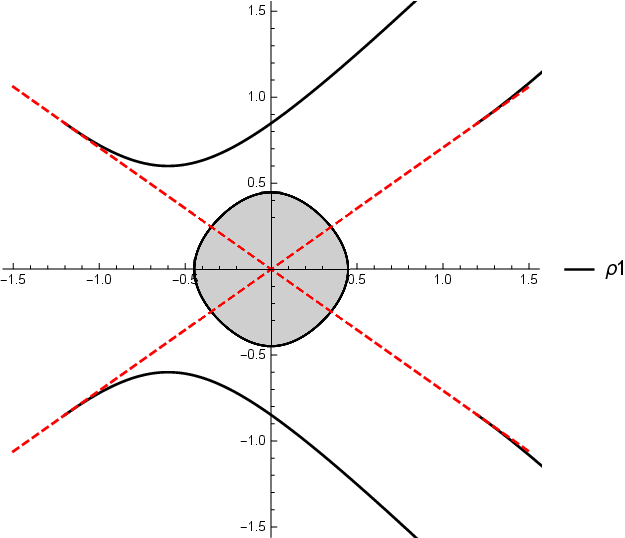} \hspace{.2cm}
\includegraphics[scale=0.5]{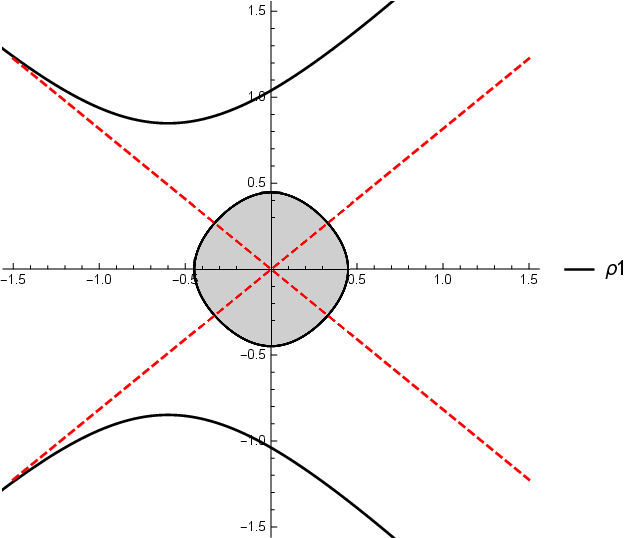}
\caption{Polar  plots for $|\rho_1|$ at fixed $\mfb=0.3$ with  $ \mfr=\mfr_0=0.424, \, \mfr=0.6 $, from left to right.}
\label{RHO1}
\end{figure}

\subsection{Comparison between the amplitudes ${\cal M}^C$ and  ${\cal M}^{NC}$ to order ${\tilde \theta}$ }

\label{IV4}

The purpose of this section is to compare both  scattering amplitudes in a suitable common region. First we identify the kinematics of the momentum-conserving case in our notation. 
In the center of mass frame we have
\barr
&& p_{1}^{\mu } =\left( \sqrt{\mathbf{p}^{2}+ {\tilde M}^{2}},\mathbf{p}%
\right) ,\quad p_{2}^{\mu }=\left( \sqrt{\mathbf{p}^{2}\mathbf{+}{\tilde M}^{2}},-%
\mathbf{p}\right) \quad 
p_{1}^{\prime \mu }=\left( \sqrt{\mathbf{p}^{\prime 2}\mathbf{+}{\tilde M}^{2}},%
\mathbf{p}^{\prime }\right) ,\quad p_{2}^{\prime \mu }=\left( \sqrt{\mathbf{p%
}^{\prime 2}\mathbf{+}{\tilde M}^{2}},-\mathbf{p}^{\prime }\right) \nonumber \\
\label{PINCONS}
\earr
with
$|\mathbf{p}^{\prime }| =|\mathbf{p}|$ and $E_{1}=E_{2}=E_{1}^{\prime }=E_{2}^{\prime }=E/2=\sqrt{|\mathbf{p}|^{2}+{\tilde M}^{2}}$.

In our notation we have 
\begin{eqnarray}
&&p_{1}^{\mu } = (E/2,\;p\sin \theta ,\;0,\;p\cos \theta ),\qquad p_{2}^{\mu } =(E/2,\;-p\sin \theta ,\;0,\;-p\cos \theta ),\nonumber  \\
&& p_{1}^{\prime \mu } =(E/2,\;p^{\prime }\cos \phi ,\;p^{\prime }\sin \phi
,\;p_{1}^{\prime }), \qquad p_{2}^{\prime \mu } =(E/2,\;-p^{\prime }\cos \phi ,\;-p^{\prime }\sin \phi
, p_{2}^{\prime })  
\label{PINCONS1}
\earr
Momentum conservation demands 
\beq
p=|\mathbf{p}|, \quad p'_1=-p'_2, \quad  |\mathbf{p}_{1}^{\prime }|^{2} =p^{\prime 2}+p_{1}^{\prime 2},
\label{MOMCONS}
\eeq
which means that in our ($ p'_1$-$p'_2$) plane, the line $X=-Y, \, (\alpha=3\pi/4)$ is selected when momentum is conserved. We define the ratio
\beq
r=\left|\frac{{\cal M}^{NC}}{{\cal M}^C}\right|_{X=-Y}
\label{RATIO}
\eeq
and calculate it for normal incidence $(\theta =0)$. In the notation where  the square bracket $[]$ denotes the dimension of the quantity inside, we have  that 
 $[\tilde \theta]=[{\tilde \lambda}]={\rm eV}$. Consequently  $[{\cal M}^{NC}]= {(\rm eV)}^{-1}$ and $[{\cal M}^{C}]={(\rm eV)}^{0}$ such that $[r]= {(\rm eV)}^{-1}$.
 
The momentum conserving amplitude is 
\beq
{\cal M}^C={\tilde \lambda}^2\left( \frac{1}{(p_{1}^{\prime }-p_{1})^{2}-m^{2}}+ \frac{1}{(p_{2}^{\prime }-p_{1})^{2}-m^{2}}   \right),
\label{MOMCONSA}
\eeq
with
\begin{eqnarray}
&&(p_{1}^{\prime }-p_{1})^{2} =2{\tilde M}^{2}-\frac{E^{2}}{2}+2 p \, p'_1, \qquad  (p_{2}^{\prime }-p_{1})^{2} =2{\tilde M}^{2}-\frac{E^{2}}{2}-2 p \, p'_1,
\end{eqnarray}%
yielding 
\begin{eqnarray}
{\cal M}^C &=&-\frac{2\tilde{\lambda}^{2}}{E^{2}}\left( \frac{2 \mathfrak{b}^2+\mathfrak{r}^2
 }{(2 \mathfrak{b}^2+\mathfrak{r}^2)^2- 4\mathfrak{b}^2 \rho^2 \cos^2 \alpha}\right),
\label{MOMCONSA1}
\end{eqnarray}%
where we  use $p'_1/E=\rho \cos \alpha$ together with $p=\sqrt{E^2/4
-{\tilde M}^2}$. 

As an intermediate case between the ultra relativistic limit and  the unrestricted value of the parameters, let  us consider  ${\tilde M} \neq 0, \, m=0=\mathfrak{r} $, which describes the exchange of a massless $\theta$-particle analogous to the photon. Here we have
\begin{eqnarray}
\left|{\cal M}^{C} \right|_{X=-Y}&=&-\frac{2 \tilde{\lambda}^{2}}{E^{2}}\left( \frac{1 }{2 \mathfrak{b}^2 -
\;{\rho ^{2}} }\right).
\label{MOMCONSA11}
\end{eqnarray}%

Next we start from  the general form of the non-conserving amplitude to order ${\tilde \theta}$
\beq
\mathcal{M}^{NC}=\tilde{\lambda}^{2}\tilde {\theta}\left( \frac{1}{(q_{1}\cdot
q_{1}-m^{2})(q_{2}\cdot q_{2}-m^{2})}+\frac{1}{(\tilde{q}_{1}\cdot \tilde{q}%
_{1}-m^{2})(\tilde{q}_{2}\cdot \tilde{q}_{2}-m^{2})}\right),  
\eeq
together with the denominators evaluated  at $X=-Y$ and $m=0$
\begin{eqnarray}
&&\left( q_{1}\cdot q_{1}-m^{2}\right) =  \left( q_{2}\cdot q_{2}-m^{2}\right)= -4 \mathfrak{b}(2 \mathfrak{b}+\sqrt{2} \rho),   \nonumber \\
&&\left( \tilde{q}_{1}\cdot \tilde{q}_{1}-m^{2}\right) =\left( \tilde{q}_{2}\cdot \tilde{q}_{2}-m^{2}\right)=  -4 \mathfrak{b}(2 \mathfrak{b}-\sqrt{2} \rho)   ,   
\end{eqnarray}%
yielding 
\beq
\left|{\cal M}^{N C} \right|_{X=-Y}=\frac{{\tilde \lambda}^2 {\tilde \theta}}{E^4 \mathfrak{b}^2 } \left(\frac{2 \mathfrak{b}^2+ \rho^2}{(2 \mathfrak{b}^2- \rho^2)^2} \right)    .
\eeq 
The final result is 
\begin{equation}
r=\left|  \frac{{\cal M}^{NC}}{{\cal M}^{C}} \right|_{X=-Y} =\frac{\tilde \theta}{2 E^2 \mathfrak{b}^2} \left(\frac{2 \mathfrak{b}^2+ \rho^2}{2 \mathfrak{b}^2- \rho^2} \right)
\label{RATIOM} 
\end{equation}
Though this ratio is suppressed by the factor ${\tilde \theta}/E^2$, the presence of singularities  in the allowed region (shaded area in  the figures)  might serve as an enhancement factor when $\rho$ is close the maximum allowed value of $\sqrt{2}\mathfrak{b}$ along $X=-Y$.  The  ratio (\ref{RATIOM}) further simplifies in  the ultrarelativistic limit $E > > ({\tilde M}, m) $,  when  we have  $\mathfrak{b}= 1/2$.
The case $\mathfrak{r}\neq 0$ is more elaborate and we only make a few comments.  In the line  $X=-Y$, the amplitude  ${\cal M}^C$ diverges for $\rho=(2 \mathfrak{b}^2+\mathfrak{r}^2)/(\sqrt{2} \mathfrak{b})> \sqrt{2} \mathfrak{b} $, that is to say,  outside the allowed region for  $\rho$. Then any enhancement factor can appear only  due to the vicinity of a singularity 
in the amplitude ${\cal M}^{NC}$. These do not arise from the contribution $\rho_1$, as seen in Fig \ref{RHO11} and Fig. \ref{RHO1}. On the contrary, $\rho_3$ might produce singularities  inside the shaded region. To find them  we look for the value of $\rho_3$ in the line  $X=-Y$  obtaining 
\beq
\rho_3(\alpha=3\pi/4)= \left | \frac{\mathfrak{r}^2-2\mathfrak{b}^2}{\sqrt{2}\mathfrak{b}} \right|\equiv \rho_{30},   
\eeq
after taking the limit in the corresponding Eq.(\ref{CEROSDEN}). We distinguish two cases recalling that the allowed region occurs for $\rho<\sqrt{2} \mathfrak{b}$ along the line. When $\mathfrak{r} < \sqrt{2} \mathfrak{b}$ we have $\rho_{30}=\sqrt{2} \mathfrak{b}- \mathfrak{r}^2/(\sqrt{2} \mathfrak{b}) < \sqrt{2} \mathfrak{b}$ so that the singularity falls inside the allowed region. This can be seen in the second plot of Fig. \ref{RHO33}, for example. In the complementary case
 $\mathfrak{r} > \sqrt{2} \mathfrak{b}$ we have to demand  
 $(\mathfrak{r}^2-2\mathfrak{b}^2)/(\sqrt{2} \mathfrak{b})\leq  \sqrt{2} \mathfrak{b} $ for  the singularity to be  available. This yields $\mathfrak{r}\leq 2 \mathfrak{b}$. The equality is illustrated in the second plot in Fig. \ref{RHO3} where the intersection of the upper curve of the graph with the boundary of the shaded area
 must correspond to $\alpha=3\pi/4$. Notice that this is consistent with the property $\alpha_{c3} < \pi/4$. Clearly, $\rho_{30}$ falls outside the allowed region   when $ 2 \mathfrak{b} \leq \mathfrak{r} < \infty$  and no enhancement can take place.

\section{Summary and Conclusions}

\label{V}


Our results can be  summarized in two topics: (i) the quantization of the $\theta$-scalar field $\Phi$ in the presence of the interface $\Sigma$ at $z=0$ and (ii) the study of phenomena arising as  a consequence of the non-conservation of the linear momentum in the direction perpendicular to the interface. 

The quantization is simplified by   using separation of variables so that the each component of the field is a product of a standard plane wave propagating parallel to the interface in the $x$-$y$ directions, times a $z$-dependent  field that carries the information of the interface via the boundary conditions. In an abuse of language w e refer to the latter as the normal modes of the system, which trivially combine with the contribution in the orthogonal directions. We label them by $\phi^k(z)$ where $k$ is always positive.
The normal modes are a triplet of plane waves in the $z$ axis resulting from the application of the boundary conditions at the interface. They are labeled as  ingoing modes when we have only an incident wave impinging on the interface. This means we can have either left  (L) or  right (R) ingoing modes as shown in Fig. \ref{ingoing}. These modes represent a particle entering the interface from either side. An explicit form of these modes  is given in Eqs. (\ref{LMODE}) and (\ref{RMODE}). To deal with particles coming out from the interface  we introduce the outgoing L-R modes composed only by one plane wave leaving the interface  as depicted in Fig. \ref{outgoing}. Outgoing modes are defined as the  complex conjugate of the ingoing modes  and  they are related also by a symmetric unitary transformation as shown in Eqs. (\ref{entrantes-salientes})-(\ref{SYMUNIT}). The component of the momentum $k_z$ of a particle described  by an ingoing or outgoing mode is related to the label $k$ of the mode according to the Table \ref{TABLE1}.

The non-trivial calculations showing the orthogonality and completeness of the normal modes are relegated to the Supplementary Materials \cite{SuppMaterials}, presenting  only the results in the main text.
The quantization follows the standard procedure of writing the field $\Phi$ either as a superposition of $L$-$R$ ingoing modes or $L$-$R$ outgoing modes, supplemented by the plane waves in the orthogonal directions,   and introducing the corresponding creation and annihilation operators according to Eqs.  (\ref{FIELD}) and (\ref{FIELD_DET_MODES}), respectively. In parallel  we calculate the Hamiltonian which we impose to  be positive definite for a general field configuration by demanding ${\tilde \theta}$ to be negative. The same condition is necessary to have the completeness  of the normal modes. We interpret this apparent restriction as a consequence of the liberty we have in choosing the sign of the coupling $\theta^\alpha\Phi \partial_\alpha \Phi$
in our initial Lagrangian (\ref{LAG1}).  In other words, given a configuration  as in Fig. \ref{system}, the sign of ${\tilde \theta}$ is fixed. When ${\tilde \theta}$ is positive (negative) we must choose the plus (minus) sign in from of the $\theta$-coupling to have a positive definite Hamiltonian. In terms of the creation and annihilation operators the Hamiltonian and the total momentum in the direction perpendicular to the interface $ {\mathbf P}_\perp$ reduce to the expected expressions (\ref{Hamiltoniano1}) and (\ref{PPERP}), respectively. As a final step in the quantization we calculate the Feynman propagator of the field $\Phi$, which is written in the convenient form (\ref{propagador2})  
showing explicitly the momentum violation contribution as well as the correct limit when ${\tilde \theta}=0$.

The next applications we deal with focus upon the momentum non-conserving effects produced  by the interface. First we consider a standard scalar particle $\Psi $ with mass $M$, which does not feel the interface,  and calculate its decay rate in the center of mass into two $\theta$-scalar particles. The interaction is given by the  Lagrangian (\ref{LAGPSIPHI}). Momentum non conservation of the outgoing particles open new channels in the  decay such as the outgoing $L$-$L$ and $R$-$R$ modes whose widths are calculated  in Eq. (\ref{DSS}). The decay threshold is also modified and we have the following result: when $4m^2 < M^2< 4m^2+ \pi^2{\tilde \theta}^2/4$ only the $L$-$L$ and $R$-$R$ modes are open, while $4m^2+ \pi^2{\tilde \theta}^2/4 <M^2$ allows the appearance of the outgoing $L$-$R$ mode. For the total decay rate $\Gamma_T$ we find $\Gamma_0 <\Gamma_T < 1.64 \Gamma_0$ where $\Gamma_0$ is the decay rate with no interface. 

In our second application we consider the scattering of standard charged scalar particles $\chi^\pm$ mediated by a $\theta$-scalar particle, through the interaction ${\tilde \lambda} \Phi \chi^\dagger \chi$. We take two incoming equal charges in their center of mass going into a two-body final state, such that only the $t$ and $u$ channels contribute. Our main interest is in the momentum-violation contribution to the scattering amplitude whose final expression in Eq. (\ref{NEWAMP}) lacks the delta function corresponding to the momentum conservation in the $z$-direction perpendicular to the interface. The leaves three undetermined variables in the final state, instead of the usual two when  we have  full four momentum conservation.
The  kinematics is described in section \ref{IV1} and the free final variables are chosen to be  the $z$ component of the momentum of each outgoing particle, $p'_1$ and $p'_2$,  together with the angle $\phi$ measuring the rotation of the projection of the center of mass direction on the interface, where the perpendicular momentum is conserved.  We obtain the scattering amplitude in section \ref{IV2} and select the momentum non-conserving contribution  ${\cal M}^{NC}$ in Eq. (\ref{T_theta1}). Our aim is to determine a measure for the non-zero probability that allows these events to occur. We obtain the  transition probability per unit time $d {\dot P}$ in Eq. (\ref{TOTPROB11}) and we introduce the concept of flux per unit length $\tilde F$ in (\ref{NEWFLUX}), which serves to define the following  generalization of the standard cross section in the center of mass
$$
d \,\Xi=\frac{d {\dot P}}{\tilde F}= \frac{1}{32 \pi^3}
\frac{1}{|{\mathbf p}|E^2} \, \left|{\cal M}^{NC}\right|^{2}
\, dp'_1 dp'_2 \, d \phi.
$$
The quantity $\Xi$, with dimensions of ${\rm cm}^3$ measures the abundance  of momentum-violating events in the scattering  process. 

Next we focus on the scattering amplitude ${\cal M}^{NC}$ and determine the allowed kinematical region in the $p'_1$-$p'_2$ plane, together with the location of its poles. At this stage we restrict ourselves to normal incidence ($\theta =0$) and introduce the convenient polar notation in (\ref{DEFPARAM}) for the $p'_1$-$p'_2$ plane. The allowed region is determined by the condition $0<\rho < \rho_-$, with  $\rho_-$ given in the second Eq. (\ref{RHOPM}) and it is depicted  as a shaded area in all the included polar plots. There we also indicate the loci of the poles of ${\cal M}^{NC}$ determined by Eqs. (\ref{CEROSDEN}) in different approximations. Symmetry considerations  imply that it is sufficient to plot  the lines $\rho_1(\alpha)=0=\rho_3(\alpha)$  determining such poles, such that the full description  is obtained after adding  a rigid rotation of the graph  by $\pi$ radians. The simplest case is when we take $m=0 \, (\mathfrak{r}=0)$ which is plotted in Figs. \ref{FIG0}-\ref{FIG1} for different choices of $\mathfrak{b}=p/E$. In this case, the poles of the scattering amplitude fall outside of the allowed momentum-violating region, displaying  touching points only at the  angles $\alpha= (2n+1)\pi/4$, where $\rho=\sqrt{2} \mathfrak{b}$.
The case $\mathfrak{r}\neq 0$ is more elaborate since the solutions giving the position of the poles acquire imaginary contributions in some angular  ranges determined by the  parameters in Eqs. (\ref{ANGR1}) and (\ref{ANGR2}) together with the indications just before these equations. We observe that the angles determining the forbidden angular regions range from zero to $\pi/4$ over the whole definition interval.  
The forbidden angular ranges are shown by dashed lines in the figures, and they behave as asymptotes where the loci displaying   the poles collapse. One  branch of the loci determined by $\rho_3=0$ penetrates  the shaded region while the other branch always recedes from it,  as shown in Figs. \ref{RHO33}-\ref{RHO3}. In this case  we have the  critical angle $\alpha_{c3}$ in Eq. (\ref{ANGR1}), determined by $\mathfrak{r} > \sqrt{2}\,\mathfrak{b} $,  beyond which an imaginary part develops. On the other hand, the loci determined by $\rho_1$ always recedes from the shaded region and has an imaginary part for all values of $\mathfrak{r}$,  according to Eq. (\ref{ANGR2}). These features  are shown in Figs. 
\ref{RHO11}-\ref{RHO1}.  

As a final result in this work we compare the magnitudes of the momentum non-conserving scattering amplitude ${\cal M}^{NC}$ with that of the momentum conserving contribution
 ${\cal M}^{C}$ by defining its ratio $r$ in Eq. (\ref{RATIO}). Since momentum conservation sets $p'_1=-p'_2$ we choose the line $X=-Y$, in the notation of Eq. (\ref{DEFPARAM}), to set the comparison. In the case $m=\mathfrak{r}=0$  we obtain
\begin{equation}
r=\left|  \frac{{\cal M}^{NC}}{{\cal M}^{C}} \right|_{X=-Y} =\frac{\tilde \theta}{2 E^2 \mathfrak{b}^2} \left(\frac{2 \mathfrak{b}^2+ \rho^2}{2 \mathfrak{b}^2- \rho^2} \right),
\label{RATIOM1} 
\end{equation}
showing that the  ratio is suppressed by the factor ${\tilde \theta}/E^2$. Nevertheless,  the presence of singularities  in the allowed region $0< \rho < \sqrt{2} \mathfrak{b}  $ might serve as an enhancement factor when $\rho$ is close the maximum allowed value of $\sqrt{2}\mathfrak{b}$ along the line  $X=-Y$ .  The  ratio (\ref{RATIOM1}) further simplifies in  the ultrarelativistic limit $E > > ({\tilde M}, m) $,  when  we have  $\mathfrak{b}= 1/2$.

The case $\mathfrak{r}\neq 0$ is more involved and we only make a few comments.  When $X=-Y$, the amplitude  ${\cal M}^C$ diverges for $\rho=(2 \mathfrak{b}^2+\mathfrak{r}^2)/(\sqrt{2} \mathfrak{b})> \sqrt{2} \mathfrak{b} $, that is to say,  outside the shaded region for  $\rho$. Then, any enhancement factor can appear only  due to the vicinity of a singularity 
in the amplitude ${\cal M}^{NC}$. These do not arise from the contribution $\rho_1$, as seen in Fig \ref{RHO11} and Fig. \ref{RHO1}. On the contrary, $\rho_3$  produces singularities  inside the shaded region only in the range $0 < \mathfrak{r}< 2 \mathfrak{b}$, as discussed in detail at the end of section \ref{IV4}.  In this way, momentum violating events may become important for some choices in the parameter region.

\acknowledgments
The authors acknowledge support from the projects CONAHCyT (México) \#CF-428214 and DGAPA-UNAM-IG100224. \\

\textbf{Data Availability Statement}: The authors declare that the data supporting the findings of this study are available within the paper and its supplementary information files.

\end{document}